\documentclass[aps,superscriptaddress,letter,nopacs,nofootinbib,longbibliography]{revtex4-2}

\usepackage{graphicx,epsfig}
\usepackage[caption=false]{subfig}
\usepackage{amsmath}
\usepackage{amsfonts}
\usepackage{fancyhdr}
\usepackage{siunitx}
\usepackage{tikz}
\usepackage{appendix}
\usepackage{hyperref}
\usepackage[compat=1.1.0]{tikz-feynman}
\tikzfeynmanset{warn luatex=false}
\usepackage{dsfont}
\usepackage{multirow}

\usepackage{comment}

\newcommand{\bsc}{Barcelona Supercomputing Center (BSC)}
\newcommand{\UniZar}{Departamento de Física Teórica, Centro de Astropartículas y Física de Altas Energías (CAPA), Universidad de Zaragoza, 50009 Zaragoza, Spain.}
\newcommand{\CQT}{Centre for Quantum Technologies, National University of Singapore, Singapore.}
\newcommand{\ICC}{Institut Cartogràfic i Geològic de Catalunya, Passeig de Santa Madrona, 45, 08038 Barcelona, Spain.}

\begin{document}

\title{Gauge and diffeomorphism invariance from quantum information principles}

\author{Claudia Núñez}
\affiliation{\ICC}
\author{Miguel Pardina}
\affiliation{\UniZar}
\author{Manuel Asorey}
\affiliation{\UniZar}
\author{José Ignacio Latorre}
\affiliation{\CQT}
\author{Alba Cervera-Lierta*}
\affiliation{\bsc}
\email{alba.cervera@bsc.es}

\begin{abstract}
Entanglement is a hallmark of quantum theory, yet it alone does not capture the full extent of quantum complexity: some highly entangled states can still be classically simulated. Non-classical behavior also requires magic, the non-Clifford component that enables universal quantum computation. Here, we investigate whether the interplay between entanglement and magic-state resources constrains the structure of fundamental interactions. 
We study gluon-gluon and graviton-graviton scattering at tree level. We focus on the high-energy limit, where mass-dependent terms are negligible, and conformal symmetry is preserved. In this regime, all particles behave as massless degrees of freedom, allowing to isolate their transverse helicities as two-qubit states. We explicitly break gauge and general covariance by modifying the quartic vertices and analyzing the resulting generation of entanglement and magic. We find that imposing maximal entanglement (MaxEnt) alone does not uniquely recover gauge-invariant and diffeomorphism invariant interactions, but adding the condition of minimal, but nonzero, magic-state generation singles it out. Our results indicate that nature favors MaxEnt and low magic: maximal quantum correlations with limited non-Cliffordness, sufficient for universal quantum computing but close to classical simulability. This dual informational principle may underlie the emergence of gauge invariance in fundamental physics.
\end{abstract}

\maketitle

\section{Introduction}

Entanglement is often regarded as the core feature of quantum theory, capturing correlations that cannot be explained classically \cite{acin2002quantum,acin2012randomness}. Yet, entanglement alone does not fully characterize quantum complexity: highly entangled systems can still be efficiently simulated on a classical computer \cite{gottesman1998heisenberg, aaronson2004improved}. What distinguishes truly quantum behavior is the presence of magic, the non-Clifford component of a quantum state or process, that enables universal quantum computation and resists classical description. In this sense, while entanglement is necessary to go beyond local realism, magic-state resources are required to go beyond classical simulability \cite{iannotti2025entanglement}. 

These observations raise a fundamental question: \textit{are the interactions governing elementary particles structured so that they both generate entanglement and escape classical simulability?} In other words, could the underlying principles of nature be constrained not only by symmetry requirements, such as gauge invariance, but also by quantum information–theoretical principles such as entanglement and magic state generation?

Recent developments support this viewpoint. By imposing maximal entanglement (MaxEnt) generation, quantum electrodynamics (QED) at tree-level can be recovered \cite{CerveraLierta2017maximal}, and the weak mixing angle obtained aligns closely with experimental values \cite{CerveraLierta2017maximal, Morales2024tripartite}. Similarly, enforcing MaxEnt in gluon scattering at tree-level reproduces quantum chromodynamics (QCD) and filters almost all unphysical interactions \cite{nunez2025universality}. These phenomenological works are becoming relevant as experimental progress has also made it possible to detect entanglement and non-stabilizerness at the LHC \cite{ATLASCollaboration2024observation, CMSCollaboration2024observation, CMSCollaboration2024measurements,yazgan2025measurements} or to observe Bell inequality violations in particle processes \cite{Fabbrichesi2024charmonium, Fabbrichesi2024bell, Gabrielli2024entanglement}. On the other limit, the degree of entanglement suppression itself appears to encode information about the nature of the interactions and their symmetries \cite{Beane2019entanglement, Low2021symmetry, Liu2023minimal}, the last is also being studied when imposing MaxEnt \cite{carena2025entanglement}, and it seems it can even predict the angles of the Pontecorvo-Maki-Nakagawa-Sakata matrix \cite{PhysRevD.111.056021}. The interplay between crossing symmetry and entanglement is also studied in 2 particle scattering processes in Ref.~\cite{mcginnis2025crossing}.

Intriguingly, the generation of magic states also varies among fundamental interactions. QED produces almost no magic-state resources \cite{liu2025qed}, while top quarks can generate significant amount of non-cliffordness depending on kinematics \cite{white2024magic}. By contrast, gluon-gluon and graviton-graviton scatterings exhibit very low magic-state resources at tree level \cite{gargalionis2025spin}. Other studies show that magic-state resources can identify the confinement-deconfinement transition in two-dimensional $\mathbb{Z}_{2}$ lattice gauge theories \cite{PRXQuantum.4.040317}. Furthermore, the weak mixing angle has been determined to $\sin^2\theta_W \simeq 0.231$, very close to the experimental value, under the principle of minimal magic-state generation \cite{liu2025quantum}. Collectively, these results suggest that quantum information principles, particularly those quantifying entanglement and non-Cliffordness, may play a crucial role in the structure of known interactions and in the search for physics beyond the Standard Model \cite{Aoude2022quantum, Severi2023quantum, Aoude2023probing, Bernal2023entanglement, Fabbrichesi2023stringent, Maltoni2024quantum, Maltoni2024tops, Carena2023entanglement, Kowalska2024entanglement}.

In this work, we explore whether gauge invariance, a cornerstone of modern field theory, can be recovered purely from quantum information principles. To this end, we study gluon-gluon and graviton-graviton scattering at tree level. As massless bosons, we take the two polarization states, left $L$ and right $R$, as the degrees of freedom, therefore restricting the quantum information analysis to a two-qubit pure state. As a consequence, both entanglement and magic-state resources are well-defined monotones. We explicitly break gauge invariance in QCD and general covariance in perturbative quantum gravity by modifying the four-vertex interaction while leaving untouched the kinetic term in the Lagrangian and studying how such deformations affect entanglement and magic state generation. Our analysis shows that imposing MaxEnt alone does not uniquely recover the gauge-invariant interaction, but when we further impose minimal (but nonzero) magic state generation, the physical gauge-invariant solution is uniquely singled out.

These findings suggest that nature may indeed be guided by a dual informational principle: it favors maximal quantum correlation (MaxEnt) while maintaining low, yet nonzero, magic—enough to enable universality in quantum dynamics, but limited to preserve near-classical simulability. This interplay between entanglement and magic-state resources may thus provide a new lens through which to view the origin and structure of fundamental interactions.

\section{Entanglement and magic-state resources in two-particle processes}\label{sec:background}

Our ultimate goal is to study under which circumstances non-locality and non-classical simulability emerge from fundamental interactions. In this work, we restrict our analysis to two-particle processes, thereby limiting it to tree-level processes. Both gluons and gravitons are massless bosons, which impose two polarization states, $|L\rangle$ and $|R\rangle$. Therefore, their quantum state can be modeled by a pure two-qubit state whose amplitudes come from the scattering amplitudes computed at tree-level:
\begin{equation}
    |\psi_{f}\rangle \sim \mathcal{M}_{\psi_{i}\rightarrow RR}|RR\rangle + \mathcal{M}_{\psi_{i} \rightarrow RL}|RL\rangle 
    + \mathcal{M}_{\psi_{i} \rightarrow LR}|LR \rangle + \mathcal{M}_{\psi_{i} \rightarrow LL}|LL\rangle,
    \label{eq:finalstate}
\end{equation}
where $\psi_{i}$ is the initial state and $\mathcal{M}_{\psi_{i}\rightarrow AB}$ is the tree-level amplitude for the outgoing particles having $AB$ polarization $A,B\in\{R,L\}$. We dropped the ket notation for simplicity. In the high-energy regime ($\sqrt{s} \gg m$) that we are considering, the restriction to tree-level scattering and massless degrees of freedom is natural. At these fundamental scales, mass-dependent terms are negligible and longitudinal polarization states become dynamically irrelevant, which reduces the asymptotic states to pure two-qubit systems defined uniquely by their transverse helicities, $|L\rangle$ and $|R\rangle$. Also, in this ultra-relativistic limit, the fundamental interactions exhibit conformal symmetry and therefore tree-level amplitudes serve as a direct candidate to check the gauge and diffeomorphism structure of the theories.

The reduction of the problem to two-qubit pure states substantially simplifies the entanglement and non-locality analysis. Any two-qubit entangled state that is pure violates a Bell inequality \cite{horodecki1995violating}. Moreover, all entanglement measures are equivalent to each other for two-qubit pure states \cite{Nielsen1991conditions}. As a consequence, we can choose a convenient figure of merit to quantify the amount of entanglement generated and the result will also imply the amount of non-locality of that state. Taking a general two-qubit pure state,
\begin{equation}
     |\psi\rangle =  \alpha|RR\rangle + \beta|RL\rangle + \gamma|LR\rangle + \delta|LL\rangle,
    \label{eq:psi_general}
\end{equation}
with $\alpha, \beta, \gamma, \delta \in \mathbb{C}$ and $|\alpha|^{2}+|\beta|^{2}+|\gamma|^{2}+|\delta|^{2} = 1$, we quantify the entanglement using the concurrence,
\begin{equation}
    \Delta = 2|\alpha\delta - \beta\gamma|,
    \label{eq:delta}
\end{equation}
where $0 \leq \Delta \leq 1$. The states with $\Delta = 0$ correspond to product states and the ones with $\Delta = 1$ to MaxEnt states.

Systems that generate high entanglement are not necessarily hard to simulate. Indeed, one can construct highly entangled states (even Absolutely Maximally Entangled States \cite{helwig2013absolutely,Cervera2019quantum}) in a quantum computer using only Hadamard and Controlled-Z gates \cite{hein2004multiparty} and still be able to represent that state with classical resources efficiently \cite{gottesman1998heisenberg}. One needs to use a quantum gate outside the Clifford group to make things hard to reproduce by a classical computer, i.e. to perform universal quantum computation. This property is typically exemplified by the number of $T$ gates (non-Clifford gates) necessary to construct such states. In other words, the amount of non-Cliffordness in a quantum state is a measure of how hard it is to simulate it with classical resources or how much power does that state have to be used as a resource for quantum information protocols.

Magic-state resources are used to quantify the amount of non-Cliffordness. 
As happens with entanglement measures, magic-state measures are diverse and not always well-defined for different systems \cite{liu2022many}. Again, for pure states this discussion simplifies and several magic measures are equivalent to each other. For that reason, we conveniently chose the Stabilizer Renyi Entropies (SRE) \cite{leone2022stabilizer, Haug2023stabilizerentropies} that are easier to compute than other measures. They are defined as
\begin{equation}
    M_{\alpha}\left(|\psi\rangle\right)=\frac{1}{1-\alpha}\log\left(\frac{1}{4}\sum_{P\in \mathcal{P}_{n}}|\langle\psi|P|\psi\rangle|^{2\alpha}\right),
    \label{eq:Malpha}
\end{equation}
where $\mathcal{P}_{n}$ is the $n-$qubit family of Pauli string operators, which for the $n=2$ is $P_{2} = P_{i} \otimes P_{j},\;\; P_{i}, P_{j} \in \{\mathds{1}, \sigma_{x}, \sigma_{y},\sigma_{z}\}$, such that a Pauli or 2 × 2 identity matrix acts on each individual qubit. Then, for 2-qubits we have 16 different Pauli string operators.
Notice that a MaxEnt state of two qubits may have $M_{\alpha}=0$ (e.g. a Bell state), while one can find a product state that has $M_{\alpha}>0$ (e.g. the $T\otimes T|00\rangle$ state). Therefore, classical simulability is linked to both high entanglement and magic.

\section{Entanglement and magic-state generation in gluon and graviton scattering}\label{sec:generation_entanglement}

In this section, we review the entanglement and magic-state properties of gluons and gravitons scattering at tree-level \cite{nunez2025universality,gargalionis2025spin} in terms of the polarization states $R$ and $L$. This process involves four Feynman diagrams, the $s$, $t$ and $u$ and quartic channels. The total scattering amplitude becomes (see Methods secton for details):
\begin{equation}
    \mathcal{M} = \mathcal{M}_{s} + \mathcal{M}_{t} + \mathcal{M}_{u} + \mathcal{M}_{4}.
    \label{eq:M_channels}
\end{equation}
We are interested only in the generation of entanglement and magic, not in its possible transformation or conservation. Therefore, we restrict the initial state to be a product state of the polarizations.

To analyze entanglement, we us the concurrence $\Delta$ as the figure of merit. For gluons, entanglement is only generated when the initial polarizations are opposite:
\begin{equation}
    \Delta_{RL}^{\textrm{gluons}} =   \frac{2 t^{2}u^{2}}{t^{4} + u^{4}},
    \label{eq:cocurrenceMandelstam}
\end{equation}
and similarly for initial $LR$. MaxEnt is achieved when $t$ and $u$ channels are indistinguishable, i.e. the center of mass frame angle is $\theta_{COM}=\pi/2$ and independent of the color charge. Remarkably, any channel can generate some amount of entanglement by its own, but all of them are necessary to achieve the maximum.

To quantify the magic, we use the stabilizer Renyi entropy of degree 2 $M_{2}$ (see Methods sections for details). The magic-state resources in gluon scattering is only non-zero when the initial polarizations are opposite. It takes the form
\begin{equation}
    M_2(|\psi\rangle_{RL})=-\log \left(\frac{t^{16}+14 t^{8} u^{8}+u^{16}}{\left(t^4+u^4\right)^4}\right) \; . \label{eq:magicYM}
\end{equation}
Its maximum value is achieved at $M_{2}^{max} = \log(4/3) \sim 0.288$ when for a center of mass frame angle $\theta_{COM} \sim 1.35$.

Similarly for gravitons, only opposite polarizations can generate entanglement,
\begin{equation}
    \Delta_{RL}^{\textrm{grav.}} =   \frac{2 t^{4}u^{4}}{t^{8} + u^{8}}.
    \label{eq:cocurrenceMandelstamGrav}
\end{equation}
and the maximum is achieved when $\theta_{COM} = \frac{\pi}{2}$, i.e. $t = u$. The fact that Eq.~\eqref{eq:cocurrenceMandelstamGrav} looks like the ``squared'' version of the Eq.~\eqref{eq:cocurrenceMandelstam} is expected from the KLT relations \cite{Kawai:1985xq}.

Likewise, the magic-state resources in graviton scattering becomes 
\begin{equation}
    M_2(|\psi\rangle_{RL})=-\log \left(\frac{t^{32}+14 t^{16} u^{16}+u^{32}}{\left(t^8+u^8\right)^4}\right) \; . \label{eq:magicGravity}
\end{equation}
which achieve its maximum value of $M_{2}^{max}= \log(4/3)\sim 0.288$ for $\theta_{COM}\sim1.46$, same $M_{2}$ as gluons. For details, see App.\ref{app:amplitudesGrav} and App.\ref{app:details_gravitons}.

\section{Recovering gauge invariance from MaxEnt and minimum magic}\label{sec:MaxEntk}

We want to explore if quantum informational quantities, such as entanglement and magic, impose certain structures in fundamental interactions. However, interactions such as QCD and perturbative quantum gravity are fixed by gauge invariance. Therefore, we must explicitly break gauge invariance to discern if quantum information is at the core of such interactions as well.

There are many ways of breaking gauge invariance. We chose what we consider a minimal approach for the tree-level calculations that we are carrying out in this work. This approach consists of breaking gauge and diffeomorphism invariance in the interaction term only, i.e. we left the kinetic term untouched. This way, we avoid modifying the degrees of freedom and we can continue using only two helicity eigenstates for the gluons and gravitons. Moreover, the interaction term is modified in a way that only the 4-vertex coupling is affected, which allows us to re-use the scattering amplitudes already computed for standard graviton and gluon scattering. Phenomenologically, it implies introducing an extra 4-vertex term in the Lagrangian with a certain structure so it only affects the 4-vertex Feynman diagram by modifying the coupling by a rescaling factor $k$. Explicitly for QCD, Lagrangian density takes the form
    \begin{equation}
        \tilde{F}_{\mu\nu}^{a}\tilde{F}^{\mu\nu a}= -\frac{1}{4}F_{\mu\nu}^{a}F^{\mu\nu a} -\frac{1}{4}\lambda_{QCD} \Lambda ^{abcd}A^{a}_{\mu} A^{b\mu} A^{c}_{\nu} A^{d\nu},\\
    \end{equation}
where $F_{\mu\nu}^{a}= \partial_{\mu} A^{a}_{\nu} - \partial_{\nu} A^{a}_{\mu} + g f_{bc}^{a}A^{b}_{\mu} A^{c}_{\nu}$. Moreover, we chose the same color dependence as the standard QCD lagrangian, $\Lambda ^{abcd}=\Lambda_{QCD}^{abcd}$. Taking $\lambda_{QCD}\equiv (k-1)g^2$ only changes the 4-vertex Feynman rule, which is now proportional to $-ikg^2$ instead of $-ig^2$. Therefore, for $k=1$ we recover the QCD gauge-invariant interaction. In perturbative quantum gravity, this breaking in the total interaction Lagrangian density takes the form
    \begin{equation}
        \tilde{\mathcal{L}}=\mathcal{L}_{2} + \mathcal{L}_{3} + \mathcal{L}_{4} + \lambda_{GR} \mathcal{L}_{4},
    \end{equation}
where $\mathcal{L}_{2}, \ \mathcal{L}_{3}, \ \mathcal{L}_{4}$ are the standard 2, 3 and 4-point interactions given by the Einstein-Hilbert Lagrangian density ${2}/{\kappa^2} \sqrt{-g}\: R$ (see App.\ref{app:amplitudesGrav} and App.\ref{app:details_gravitons}). We can choose $\lambda_{GR}\equiv (k-1)$ so the only Feynman rule at tree-level modified is the coupling in the 4-vertex diagram, which becomes $-i k \kappa^2$. 
As a result, the total scattering amplitudes from Eq.\eqref{eq:M_channels} take now the form:
\begin{equation}
    \mathcal{M} = \mathcal{M}_{s} + \mathcal{M}_{t} + \mathcal{M}_{u} + k \mathcal{M}_{4}.
    \label{eq:addingk}
\end{equation}
The amplitudes per channel in gluons can be found in Ref.\cite{nunez2025universality}, 
while for gravitons are written explicitly in App.\ref{app:amplitudesGrav} and App.\ref{app:details_gravitons}. 

\begin{figure}[t!]
    \centering
    \includegraphics[width=0.5\linewidth]{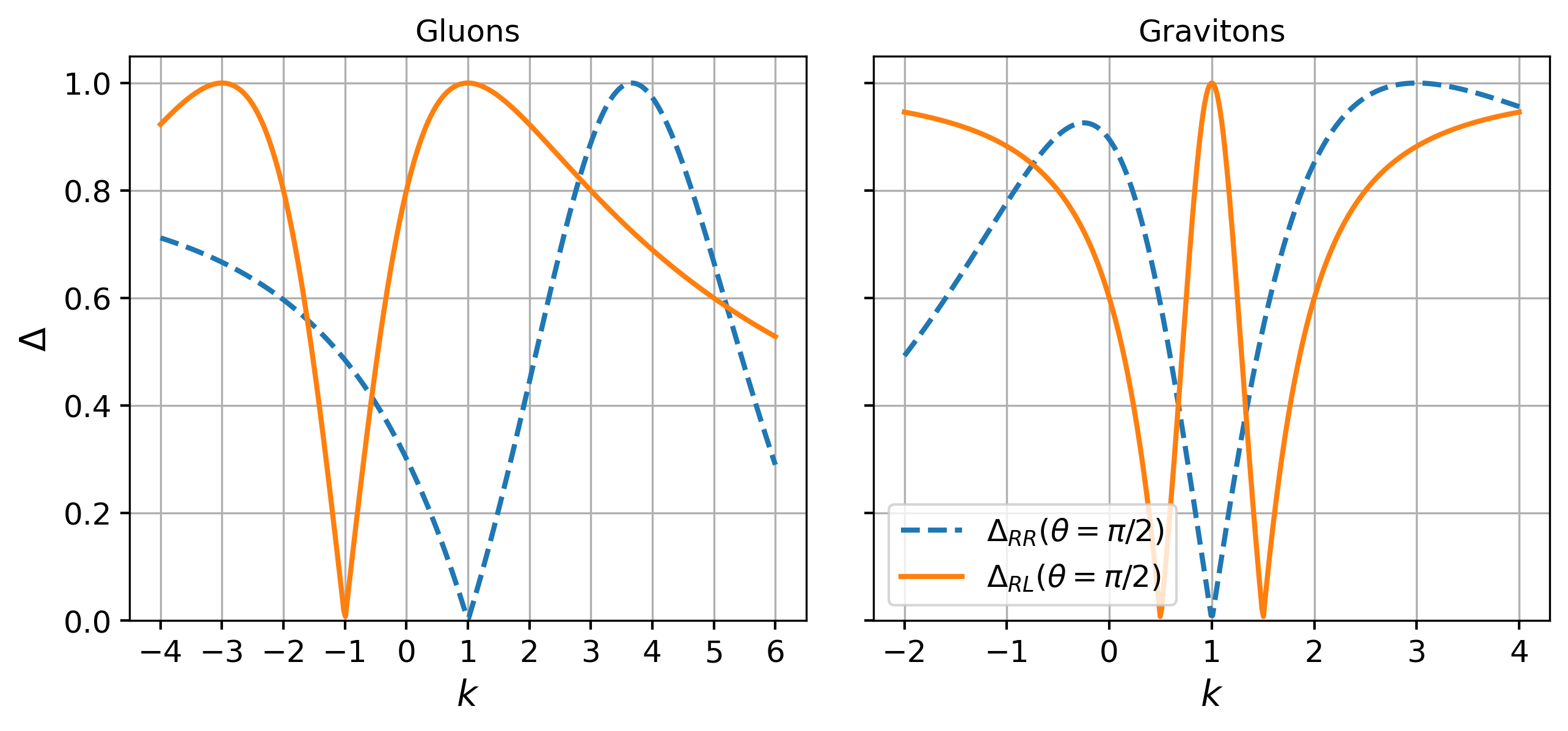}
    \caption{Concurrence as a function of the 4 vertex parameter $k$ for initial polarizations $RL$ and $RR$ and COM angle $\theta_{COM}=\pi/2$. The gauge-invariant solutions correspond to $k=1$, where MaxEnt is achieved for initial $RL$ polarization while $RR$ generate a product state. Left: for QCD, MaxEnt is also obtained for two other unphysical solutions, $k=-3$ and $k=11/3$, for an initial polarization of $|RR\rangle$. Right: in gravity, another non-physical solution is obtained, $k=3$, for an initial polarization of $|RR\rangle$. }
    \label{fig:concurrencek}
\end{figure}

If we impose MaxEnt, we recover the known gauge-invariant result for $k=1$ both for gluons and gravitons, but we also obtain other non-physical solutions. In gluons and $\theta_{COM}=\pi/2$, the concurrence becomes:

\begin{eqnarray}
    \Delta_{RL}^{gluons}(k)  &=&  \left|\frac{4 (k +1)}{5 + 2k + k^{2}}\right|, \\
    \Delta_{RR}^{gluons}(k) &=& 2 \left| \frac{2 (k -1)(k-7)}{93-34k+5k^{2}}\right| .
\end{eqnarray}
MaxEnt can also be obtained for $k=-3$ for initial $RL$ and for $k=11/3$ for initial $RR$ polarizations. Notice that these non-physical solutions imply non-zero entanglement in the other polarization scheme, while the gauge-invariant solution achieves MaxEnt in $RL$ only when we do not have entanglement at all in $RR$. We plot this result in the left plot from Fig.\ref{fig:concurrencek} for $\theta_{COM}=\pi/2$, which is independent on the color of the gluons. For other values of $\theta_{COM}$, concurrence does depend on the color, but MaxEnt is achieved for all possible color configurations for different $k$ (see App.\ref{app:details_gluons} for details).

\begin{figure}[t]
    \centering
        \includegraphics[width=0.5\linewidth]{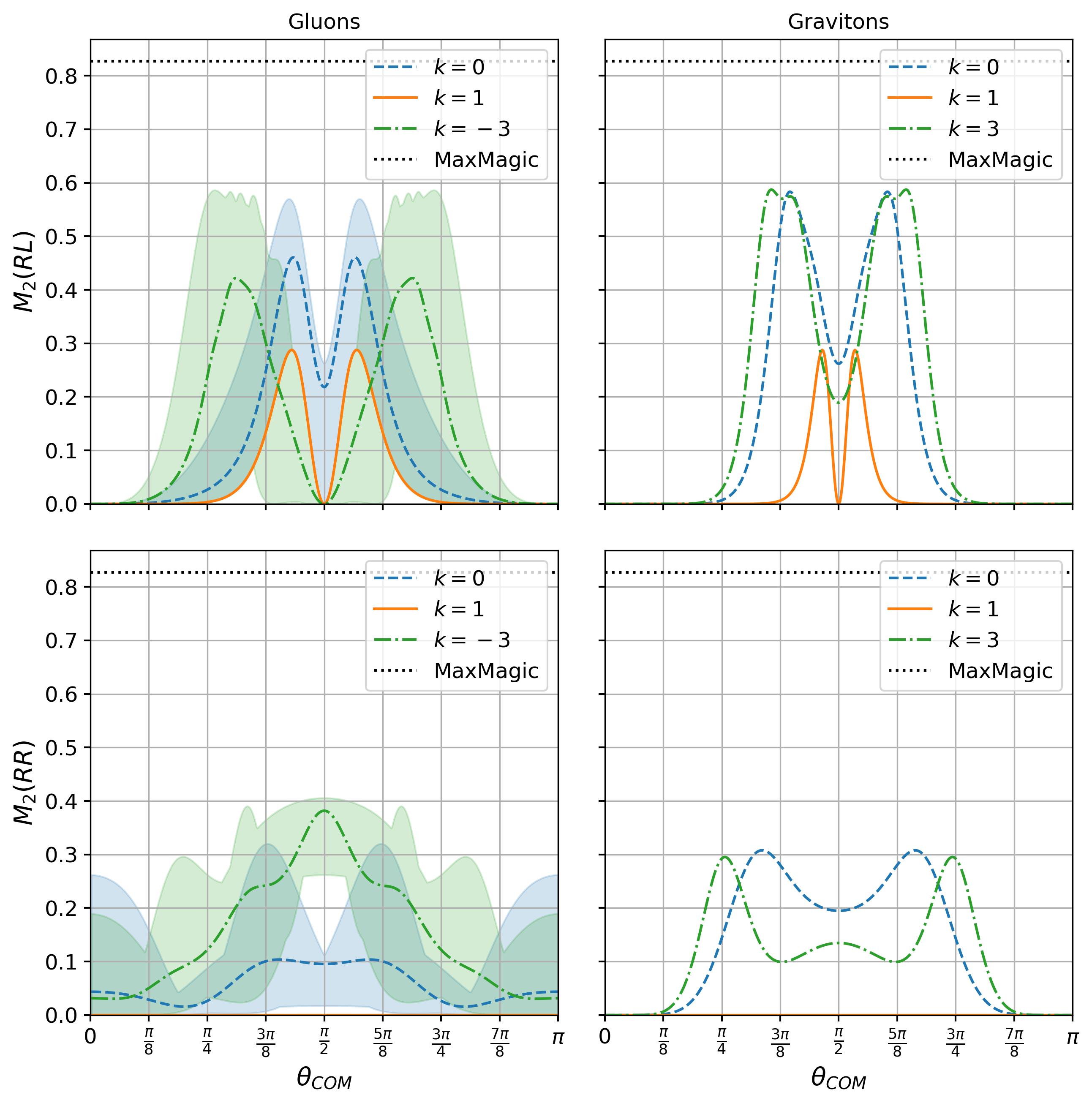}%
    \caption{Magic-state resources measured with $M_{2}$ as a function of $\theta_{COM}$ for different values of the 4-vertex parameter $k$ and initial $RL$ polarization. Left, for gluons, and right, for gravitons. For initial $RL$ polarization, $M_{2}$ has a local minima at $\theta_{COM}=\pi/2$. Then, it shows two symmetric maxima. The value of that maximum depends on $k$ and it is minimal at $k=1$, as shown Fig.\ref{fig:magick}. For initial $RR$ polarization, other values of $k$ show some magic, while $k=1$ is a product state with no magic. For $k\neq1$, the gluons result depends on the color, in particular to the relations between the structure constants $F_1\equiv f^{aa'c}f^{bb'c}$, $F_2\equiv f^{ab'c}f^{ba'c}$ and $F_{3}=f^{abc}f^{a'b'c}$: there are six possible combinations between these $F_{1}$, $F_{2}$ and $F_{3}$ specified in the plots with a shade between the maximum and minimum values, while the solid line corresponds to the median. For comparison, the maximal amount of $M_{2}$ (MaxMagic) for a pure two-qubit state is $\log(16/7)$.}
    \label{fig:magictheta}
\end{figure}

\begin{figure}[ht]
    \centering
        \includegraphics[width=0.4\linewidth]{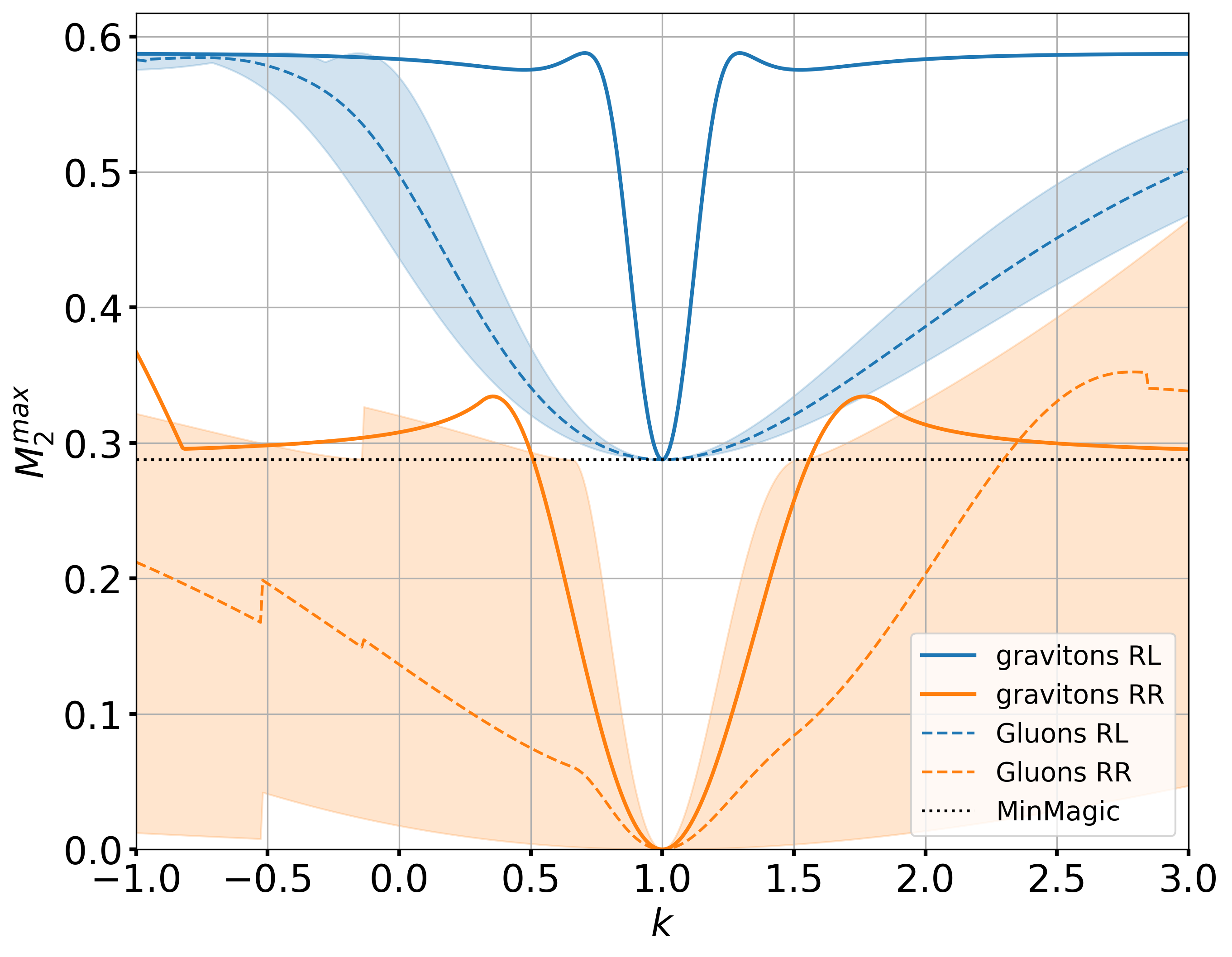}%
    \caption{Maximum $M_{2}$ achieved as a function of the 4-vertex parameter $k$. The global minimum is obtained for $k=1$, i.e. the gauge-invariant solution, and corresponds to $M_{2}=\log(4/3)$ (MinMagic) for initial $RL$ polarization, and zero por initial $RR$ polarization.}
    \label{fig:magick}
\end{figure}

Similarly, for gravitons we recover the known result for $k=1$ (MaxEnt for initial $RL$ and zero entanglement for $RR$), but also non-physical solutions. In particular, for $\theta_{COM}=\pi/2$, the concurrence for initial $RL$ and $RR$ polarizations become
\begin{eqnarray}
    \Delta_{RL}^{grav}(k) &=&  \left|1-\frac{2}{5+4k(k-2)}\right|, \\
\Delta_{RR}^{grav}(k) &=&  \left| \frac{(k -1)(3k+17)}{19-10k+7k^{2}}\right|.
\end{eqnarray}
MaxEnt is also achieved for initial $RR$ for $k=3$ and $\theta_{COM}=\pi/2$, as shown in Fig.\ref{fig:concurrencek} right. For $k>3$, we further confirm that there always exists a $\theta_{COM}$ that achieves MaxEnt.
As happened with gluons, when entanglement is maximal in $RR$, it is not minimal, i.e. zero, for $RL$.

In conclusion, Nature chooses an extremal solution, i.e. if MaxEnt is generated in opposite polarizations, it has to be zero for the same polarization scheme.

Let's analyze the amount of magic-state resources that can be generated as a function of $k$. We restrict our initial analysis to the second order SRE, $M_{2}$. We obtain a quartic polynomial profile of $M_{2}$ as a function of  $\theta_{COM}$ with the intermediate minimum centered in different $\theta_{COM}$ values depending on $k$ and with different heights. Fig.\ref{fig:magictheta} shows that dependence for gluons and gravitons with initial $RL$ and $RR$ polarizations. For gluons, the result depends on the color, in particular, it depends on $F_1= f^{aa'c}f^{bb'c}$, $F_2= f^{ab'c}f^{ba'c}$ and $F_{3}=f^{abc}f^{a'b'c}$ coefficients, except the gauge-invariant solution $k=1$.  Since there exist up to six different $F_{1}-F_{2}-F_{3}$ configurations, we plot the median and a shade between the maximum and minimum values. Notice, however, that the gauge-invariant solution $k=1$ is the only one that $M_{2}$ does not depend on the color, following a color universality similar to the one found for MaxEnt \cite{nunez2025universality}. 

We now consider the maximal amount of magic, quantified by $M_2$, as a function of $k$. In other words, for each possible value of $k$, how far from a stabilizer state are the final states generated? Since $M_2$ varies with the center-of-mass angle, we first maximize it over the allowed kinematical configurations in order to characterize the worst-case quantum complexity associated with each deformation, i.e.,
\begin{equation}
    M_{2}^{max} \equiv \max_{\theta_{COM}} M_{2}(\theta_{COM},k).
    \label{eq:max_magic}
\end{equation}
Again, for gluons, the $M_{2}$ will depend on the color. We analyze different $M_{2}^{max}$ curves depending on the possible relation between $F_{1}-F_{2}-F_{3}$ factors as we did above. In all cases, the $M_{2}^{max}$ has a global minimum at $k=1$, as shown in Fig.\ref{fig:magick}. The non-physical solutions obtained by imposing MaxEnt are not extremal points in $M_{2}^{max}$. Here it is clearer that $k=1$ is the only solution that does not depend on the color.
In gravitons, as shown in Fig.\ref{fig:magick}, the absolute minimum is also located at $k=1$. The other unphysical solutions obtained when imposing MaxEnt do not show an extremal behaviour in terms of magic.

Both for gluons and gravitons, the global minimum is $M_{2}^{max}(k=1) = \log\left(\frac{4}{3}\right)\sim 0.288$ for initial $RL$ polarization, significantly below the maximum $M_{2}$ that can be achieved by a pure two-qubit state, $M_{2}^{max}=\log\left(16/7\right)\sim 0.827$ \cite{liu2025maximal}. However, it is not zero, meaning that both gravitons and gluons interaction can generate certain magic.

\begin{figure}[t!]
    \centering
    \includegraphics[width=0.7\linewidth]{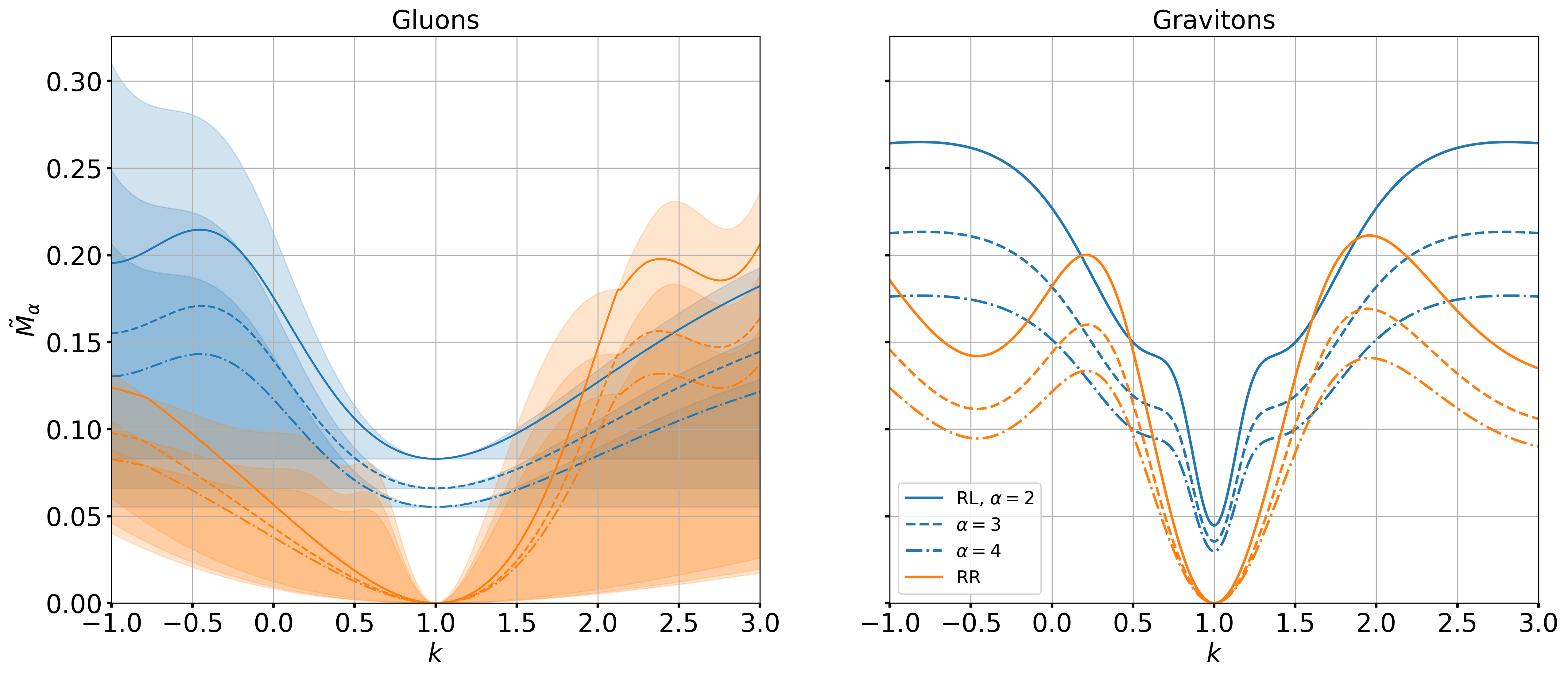}
    \caption{Stabilizer Renyi Entropies of degree $\alpha=2,3,4$ for gluons and gravitons final state scattering. The SRE are integrated over the COM angle. For gluons, the result depends on the color, so plot shows the region between maximum and minimum values and the lines correspond to the median. As happened in Fig.\ref{fig:magick}, there is a single global minimum at $k=1$, the gauge-invariant solution.}
    \label{fig:SRE_all}
\end{figure}

To analyze if this result holds for other SRE, we plot $M_{3}$ and $M_{4}$. Moreover, instead of taking the maximal $\theta_{COM}$, this time we integrate over all $\theta_{COM}$,
\begin{equation}
    \tilde{M}_{\alpha}(k)\equiv\frac{1}{2}\int_{\theta_{COM}} M_{\alpha}(k,\theta_{COM})\sin\theta_{COM}d\theta_{COM},
\end{equation}
where we introduce the $k$ and $\theta_{COM}$ dependence explicitly. Fig.\ref{fig:SRE_all} shows the result, which follows the same conclusion as Fig.\ref{fig:magick}: there is a global minimum at $k=1$, the gauge-invariant solution. Notice also that the SRE in gluons depend on the color configuration, except for the gauge-invariant point $k=1$, where all possibilities converge: the universality of entanglement in terms of the color observed in Ref.\cite{nunez2025universality} is also reproduced for SRE. In other words, the gauge invariant point is the only one where magic-state resources are independent on the color configuration, as happens with the entanglement.

\section{Conclusions}\label{sec:conclusions}

We break gauge invariance explicitly in QCD and perturbative gravity and analyze its effect in the capacity of entanglement and magic-state generation in gluon and graviton scattering at tree-level in the high-energy regime. The gauge invariant solution appears as a global minimum in terms of magic-state resources and corresponds to the generation of maximally entangled states. Our results suggest that nature favors maximal quantum correlations (MaxEnt) while maintaining low—but nonzero—magic. In this regime, interactions generate strong entanglement, yet remain close to the boundary of classical simulability. This balance appears to be a recurring feature of physical systems: maximal magic-state generation is not necessarily desirable, as excessive non-Cliffordness can hinder computational or physical efficiency. For instance, in measurement-based quantum computation (MBQC) with only Pauli measurements, too much magic-state resources leads to inefficient computation, mirroring the way excessive entanglement can be detrimental in standard MBQC architectures \cite{liu2022many}.

Nevertheless, several caveats are in order. Stabilizer Rényi entropies (SREs) provide reliable measures of magic-state resources only for pure states, and care must be taken when extending them to mixed or multipartite scenarios. The definition of a ``good” magic measure remains an active topic of research \cite{liu2022many}, with similar ambiguities known from multipartite entanglement theory. These limitations reflect the inherent complexity of quantifying nonclassical resources in realistic quantum systems. Here we only analyse qubit processes involving two particles, so there exists the possibility that one needs to go to multipartite \cite{Acin2001three-party} or higher dimensional systems to rule out certain solutions with maybe only one principle.

The rescaling of the four-gluon and four-graviton interaction terms represents a specific choice of gauge-symmetry breaking. While we regard it as a minimal deformation—since it does not modify propagators or alter the number of physical degrees of freedom beyond the quartic vertex—other forms of gauge-symmetry breaking should be investigated to determine to what extent quantum-information principles can recover gauge invariance in more general settings. 

Our magic-state resources analysis has been limited to integrating over all angles in order to study the average behavior. In principle, one could also include the probability of observing each event, i.e. weight the configurations by the differential cross section. However, once the event probability (the cross section) is taken into account, the vast majority of events would correspond to forward scattering, since we are working in the high-energy regime where forward scattering dominates. As a consequence, the most probable states would be those with $\theta \rightarrow 0$, which are product states that do not exhibit entanglement nor magic. Presenting the analysis in this way, although physically well motivated, would therefore tend to obscure the central question we aim to address: whether there exists a preference for the gauge-invariant solution. In this sense, including the full cross-section weighting would mask the underlying structure we are trying to uncover. Nevertheless, in analogy with the discussion surrounding the choice of the gauge-symmetry-breaking method, such an analysis should be incorporated in future work in order to reach more fundamental and far-reaching conclusions.

Our findings reinforce the view that the laws of nature optimize quantum informational principles. In this case, they maximize quantum correlation while minimizing non-Clifford complexity. Such a finding might be especially relevant when studying effective theories such as perturbative gravitation, where the high-energy regime may impose quantum information constraints on low energies. Understanding the role that these information-theoretic measures play in fundamental interactions can also help establish how hard it will be to simulate them using quantum devices \cite{PhysRevB.111.L081102}. Overall, this balance among entanglement, magic, and possibly other factors may represent a fundamental informational constraint that guides the structure of physical interactions.

\section*{Authors contribution}

C. N. did all the calculations involving gluons, while M. P. did all the calculations that involve gravitons. C. N., J. I. L. and A. C.-L. developed the original idea for gluons and M. P. and M. A. suggested the extension for perturbative gravity. A. C.-L. lead the work and the analysis of the results. All authors contributed to the manuscript writing and to the discussions of the results.

\section*{Acknowledgements}

M.A. and M. P. are partially supported by Spanish Grants No. PGC2022-126078NB-C21 funded by MCIN/ AEI/ 10.13039/501100011033, Diputación General de Aragón-Fondo Social Europeo (DGA-FSE) Grant No. 2020-E21-17R of the Aragon Government;
A.C.-L. acknowledges funding from Grant RYC2022-037769-I funded by MICIU /AEI /10.13039/ 501100011033 and by “ESF+".

\appendix

\section{Graviton scattering amplitudes at tree-level}\label{app:amplitudesGrav}

General relativity can be described by the well known Einstein-Hilbert action 
\begin{equation}
  S_{GR}=-\dfrac{2}{\kappa^2}\int d^4x\: \sqrt{-g}\: R \; ,
\end{equation}
where $g$ is the determinant of the spacetime metric $g_{\mu\nu}$, $R$ is the Ricci scalar, $\kappa^2=32\pi G_N$ and $G_N$ is Newton's constant. Feynman rules for Einstein's gravity are then obtained within a framework of perturbative effective gravity where the metric is decomposed as $g_{\mu\nu}=\eta_{\mu\nu}+\kappa h_{\mu\nu}$, this is, gravitational effects below the Planck scale $M_{P}^2=8\pi G_N$ are encoded in small metric perturbations $h_{\mu\nu}$ around the flat background of Minkowski spacetime $\eta_{\mu\nu}$. The perturbations $h_{\mu\nu}$ can be formally identified with the graviton, a massless particle of spin 2.

To compute the amplitudes for the four channels according to their respective Feynman rules, we employ the FeynGrav package for Mathematica \cite{Latosh:2022ydd}, which extends the capabilities of the widely used FeynCalc package in particle physics to include gravitational interactions. The kinematics and conventions used for the gravitational scattering are the same as \cite{Sannan:1986tz}. All of the following expressions are evaluated in the transverse–traceless gauge and with the on-shell condition $s+t+u=0$ imposed. The explicit complete set of scattering amplitudes for gravitons in general relativity is presented in the App.\ref{app:details_gravitons}. Here, we show its $k$-modified version:  
\begin{eqnarray}
    \mathcal{M}_{\substack{RR \rightarrow RR \\ LL \rightarrow LL}} &=& 
    \frac{i \kappa ^2 }{4}\left(\frac{s^3}{t u}+8 (k-1) \frac{t^2 u^2}{s^3}\right), \nonumber\\
    \mathcal{M}_{\substack{RL \rightarrow RL \\ LR\rightarrow LR}} &=& 
     \frac{i \kappa^2}{4}\frac{u^3}{st}, \nonumber\\
    \mathcal{M}_{\substack{RR \rightarrow LL \\ LL \rightarrow RR}} &=&  
    -2 i \kappa ^2 (k-1)\frac{ t u }{s^3}\left(2 t^2+t u+2 u^2\right), \nonumber \\
    \mathcal{M}_{\substack{RL \rightarrow LR\\LR\rightarrow RL}} &=& \frac{i \kappa^2}{4}\frac{t^3}{su}, \nonumber \\
    \mathcal{M}_{ \substack{RR \\ LL}\rightarrow  {\substack{RL \\ 
        LR}}} &=& \mathcal{M}_{ \substack{RR \\ LL}\rightarrow  {\substack{LR \\ 
        RL}}} = \mathcal{M}_{ \substack{RL \\
        LR} \rightarrow  \substack{RR \\ LL}} 
        = \mathcal{M}_{ \substack{RL \\
        LR} \rightarrow  \substack{LL \\ RR}} =
         2 i \kappa ^2 (k-1)\frac{ t^2 u^2}{s^3}.
\end{eqnarray}
Of course, this matches the sum of the total general relativity amplitudes for the case $k=1$.

\section{Entanglement in gluons}\label{app:details_gluons}

In Ref.\cite{nunez2025universality} we presented a detailed analysis of gluon scattering entanglement, both for the gauge-invariant solution and the gauge-breaking solution that depends on the four-vertex coupling $k$. However, we only discussed the $\theta_{COM}=\pi/2$ solution, but other solutions exist for other angles and $k$.
When the initial state have opposite polarizations, there always exist a $\theta_{COM}$ for some color configuration that achieves MaxEnt if $k < -1$. If it shares the same polarization, we can find MaxEnt for some $\theta_{COM}$ if $k \geq \frac{1}{3} (10 \sqrt{10}-23)\sim 2.87$ or $k = 1/3$. The range of $k$ for which MaxEnt can be achieved varies depending on the six possible relations between the structure constants for each case: $F_{1}=0$, $F_{2}=0$, $F_{1}=F_{2}$, $F_{1}=-F_{2}$,  $F_{1}=2F_{2}$ and  $2F_{1}=F_{2}$, for initial opposite polarizations, and $F_{1}=0 \neq F_{3}=-F_{2} $, $F_{2}=0 \neq F_{1}=F_{3}$, $F_{3}=0 \neq F_{1}=F_{2}$, $F_{1}=2F_{2}=2F_{3}$,  $F_{2}=2F_{1}=-2F_{3}$ and  $F_{3}=2F_{1}=-2F_{2}$, for
equal initial polarizations.

\section{Graviton scattering amplitudes at tree-level by channel}

In this section, we present the explicit amplitudes per channel of graviton scattering.

\subsection*{$s$-channel}
The non-zero scattering amplitudes of the $s$-channel take the form 
\begin{eqnarray}
    \mathcal{M}_{RR \rightarrow RR} &=& \mathcal{M}_{RR \rightarrow LL} = \mathcal{M}_{LL \rightarrow LL} = \mathcal{M}_{LL \rightarrow RR}= 
    \frac{i \kappa ^2}{4} \dfrac{(t-2u)(u-2t)}{s}.
\end{eqnarray}

\subsection*{$t$-channel}
The scattering amplitudes of the $t$-channel take the form 
\begin{eqnarray}
    \mathcal{M}_{RR \rightarrow RR} &=& \mathcal{M}_{LL \rightarrow LL} = 
    -\frac{i \kappa ^2}{4}\frac{u^2}{ s^4 t}\left(4 t^4+16 t^3 u+22 t^2 u^2+9 t u^3+u^4\right), \nonumber\\
    \mathcal{M}_{RL \rightarrow RL} &=& \mathcal{M}_{LR\rightarrow LR} = 
     -\frac{i \kappa ^2}{4}\frac{ u^4 }{s^4 t}\left(2 t^2+t u+u^2\right), \nonumber\\
    \mathcal{M}_{RR \rightarrow LL} &=& \mathcal{M}_{LL \rightarrow RR} = 
    -\frac{i \kappa ^2}{4}\frac{t^3}{s^4}\left(2 t^2+17 t u+17 u^2\right), \nonumber \\
    \mathcal{M}_{RL \rightarrow LR} &=&  \mathcal{M}_{LR\rightarrow RL} = -\frac{i \kappa ^2}{4}\frac{t^3}{s^4}\left(2 t^2+t u+u^2\right), \nonumber \\
    \mathcal{M}_{ \substack{RR \\ LL}\rightarrow  {\substack{RL \\ 
        LR}}} &=& \mathcal{M}_{ \substack{RR \\ LL}\rightarrow  {\substack{LR \\ 
        RL}}}= \mathcal{M}_{ \substack{RL \\
        LR} \rightarrow  \substack{RR \\ LL}} = \mathcal{M}_{ \substack{RL \\
        LR} \rightarrow  \substack{LL \\ RR}} =
          \frac{i \kappa ^2}{4}\frac{t^2 u^2 }{ s^4}(3 t+5 u).
\end{eqnarray}

\subsection*{$u$-channel}
The scattering amplitudes of the $u$-channel take the form 
\begin{eqnarray}
    \mathcal{M}_{RR \rightarrow RR} &=& \mathcal{M}_{LL \rightarrow LL} = 
    -\frac{i \kappa ^2}{4}\frac{t^2}{ s^4 u}\left(t^4+9 t^3 u+22 t^2 u^2+16 t u^3+4 u^4\right), \nonumber\\
    \mathcal{M}_{RL \rightarrow RL} &=& \mathcal{M}_{LR\rightarrow LR} = 
     -\frac{i \kappa ^2}{4}\frac{ u^3 }{s^4}\left(t^2+t u+2u^2\right), \nonumber\\
    \mathcal{M}_{RR \rightarrow LL} &=& \mathcal{M}_{LL \rightarrow RR} = 
    -\frac{i \kappa ^2}{4}\frac{u^3}{s^4}\left(17t^2+17 t u+2u^2\right), \nonumber \\
    \mathcal{M}_{RL \rightarrow LR} &=&  \mathcal{M}_{LR\rightarrow RL} = -\frac{i \kappa ^2}{4}\frac{t^4}{s^4u}\left(t^2+t u+2u^2\right), \nonumber \\
    \mathcal{M}_{ \substack{RR \\ LL}\rightarrow  {\substack{RL \\ 
        LR}}} &=& \mathcal{M}_{ \substack{RR \\ LL}\rightarrow  {\substack{LR \\ 
        RL}}}= \mathcal{M}_{ \substack{RL \\
        LR} \rightarrow  \substack{RR \\ LL}} = \mathcal{M}_{ \substack{RL \\
        LR} \rightarrow  \substack{LL \\ RR}} =
          \frac{i \kappa ^2}{4}\frac{t^2 u^2 }{ s^4}(5t+3u).
\end{eqnarray}

\subsection*{$4$-point}
The non-zero scattering amplitudes of the $4$-point channel take the form 
\begin{eqnarray}
    \mathcal{M}_{RR \rightarrow RR} &=& \mathcal{M}_{LL \rightarrow LL} = \mathcal{M}_{ \substack{RR \\ LL}\rightarrow  {\substack{RL \\ 
        LR}}} = \mathcal{M}_{ \substack{RR \\ LL}\rightarrow  {\substack{LR \\ 
        RL}}}= \mathcal{M}_{ \substack{RL \\
        LR} \rightarrow  \substack{RR \\ LL}} = \mathcal{M}_{ \substack{RL \\
        LR} \rightarrow  \substack{LL \\ RR}}=
   2 i \kappa ^2\frac{ t^2 u^2}{s^3}, \nonumber\\
    \mathcal{M}_{RR \rightarrow LL} &=& \mathcal{M}_{LL \rightarrow RR} = 
    -2 i \kappa ^2\frac{ t u }{s^3}\left(2 t^2+t u+2 u^2\right).
\end{eqnarray}

\subsection*{Total amplitudes for gravitons}

The non-zero total scattering amplitudes are

\begin{eqnarray}
    \mathcal{M}_{RR \rightarrow RR} &=& \mathcal{M}_{LL \rightarrow LL} = 
    \frac{i \kappa ^2 }{4}\frac{s^3}{t u}, \nonumber\\
    \mathcal{M}_{RL \rightarrow RL} &=& \mathcal{M}_{LR\rightarrow LR} = 
     \frac{i \kappa^2}{4}\frac{u^3}{st}, \nonumber\\
    \mathcal{M}_{RL \rightarrow LR} &=&  \mathcal{M}_{LR\rightarrow RL} = \frac{i \kappa^2}{4}\frac{t^3}{su}. \label{eq:totalGrav}
\end{eqnarray}

\section{Entaglement and magic-state generation in gluons}

In this section, we present the explicit quantum states generated in gluon scattering when considering the 4-vertex modification.

The normalized final states when we break gauge invariance are
\begin{align}
|\psi(k)\rangle_{RL} &= \frac{1}{\sqrt{N_{RL}}} \Big(
(F_1 + F_2)(k-1) t u \, \left(|RR\rangle + |LL\rangle \right)  \nonumber\\
&\quad+ \frac{u}{t} \left((F_1+F_2)(k-1) t u -2 F_1 s(F_2 t + u)  \right) |RL\rangle \nonumber\\
&\quad + \frac{t}{u} \left((F_1+F_2)(k-1) t u -2 F_1 s(F_2 t +  u)\right) |LR\rangle 
\Big)
\label{eq:RLYMk}
\end{align}
and
\begin{align}
|\psi(k)\rangle_{LR} &=\frac{1}{\sqrt{N_{RL}}} \Big(
(F_1 + F_2)(k-1) (t u)^2 \, \left(|RR\rangle + |LL\rangle \right)  \nonumber\\
&\quad + \frac{t}{u} \left((F_1+F_2)(k-1) t u -2 F_1 s(F_2 t +  u)\right) |RL\rangle \nonumber\\
&\quad+ \frac{u}{t} \left((F_1+F_2)(k-1) t u -2 F_1 s(F_2 t + u)\right) |LR\rangle
\Big),
\label{eq:LRYMk}
\end{align}
where
\begin{align*}
N_{RL} = N_{LR} &= \frac{1}{t^2 u^2} \Big(2 (F_1 + F_2)^2 (k-1)^2 t^4 u^4 + t^4 ((F_1 + F_2) (-1 + k) t u - 2 F_1 s (F_2 t + u))^2 \nonumber \\
& + u^4 ((F_1 + F_2) (-1 + k) t u - 2 F_1 s (F_2 t + u))^2\Big) ,
\end{align*}
for initial opposite polarizations, and
\begin{align}
|\psi(k)\rangle_{RR} &= \frac{1}{\sqrt{N_{RR}}} \Big(
\frac{1}{tu}\big(F_2 t^2 \big( 4 s^2 + 2 s (t-u) - (k-1) u (t + 2u) \big) \nonumber\\
&\quad + u \big(  F_1 u \big( 4s^2 +2s(u-t) - (k-1) t (2t+u) \big) - F_3 (k+1) st (t-u)\big)\big) |RR\rangle \nonumber \\
&\quad + (F_1 + F_2) (k-1) t u \, \left(|RL\rangle + |LR\rangle\right)
\nonumber \\
&\quad+(k-1)\big( F_3 s (t-u) - F_2 u (2 t + u) - F_1 t (t + 2 u) \big) |LL\rangle
\Big)
\label{eq:RRYMk}
\end{align}
and
\begin{align}
|\psi(k)\rangle_{LL} &= \frac{1}{\sqrt{N_{RR}}} \Big(
(k-1)\big( F_3 s (t-u) - F_2 u (2 t + u) - F_1 t (t + 2 u) \big)|RR\rangle \nonumber \\
&\quad + (F_1 + F_2) (k-1) t u \left(|RL\rangle + |LR\rangle\right)
\nonumber \\
&\quad+ \frac{1}{tu} \big(F_2 t^2\big( 4 s^2 + 2 s (t-u) - (k-1) u (t + 2u) \big) \nonumber\\
&\quad + u \big(  F_1 u \big( 4s^2 +2s(u-t) - (k-1) t (2t+u) \big) - F_3 (k+1) st (t-u)\big)\big)|LL\rangle
\Big)
\label{eq:LLYMk}
\end{align}
where
\begin{align*}
N_{RR}=N_{LL} &= \frac{1}{t^2u^2}
\Big(2 (F_1+F_2)^2 (k-1)^2 t^4 u^4 
+ (k-1)^2 t^2u^2\Big( F_3 s (u-t) + F_2 u (2 t + u) + F_1 t (t + 2 u) \Big)^2 \\
&\quad+ \Big( (k-1) tu (F_3 s (t-u) + F_1 u (2 t + u) + F_2 t (t + 2 u)) \\
&\quad+ 2s (F_2 t^2 (2s+t-u) + u (F_3 t (u-t) + F_1 u (2s-t+u)\Big)^2\Big),
\end{align*}
for equal initial polarizations.

Then, the $\alpha$-order magic-state resource for state~\eqref{eq:RLYMk} or~\eqref{eq:LRYMk} takes the form 
\begin{align}
M_{\alpha}\big(|\psi(k)\rangle_{RL}\big) & = \frac{1}{1-\alpha}\log \Bigg[\frac{1}{4}\Bigg(1 +2 \left(\frac{(t^4-u^4)E^2}{D}\right)^{2\alpha} + \left(\frac{(t^4+u^4)E^2- 2t^4u^4(F_{1} + F_{2})^2(k-1)^2}{D} \right)^{2\alpha} \nonumber\\
& + 2^{1+2\alpha}\left(\frac{(F_{1} + F_{2})(k-1)t^2u^2E(t^2+u^2)}{D}\right)^{2\alpha} + 2^{1+2\alpha}\left(\frac{(F_{1} + F_{2})(k-1)t^2u^2E(t^2-u^2)}{D}\right)^{2\alpha}\nonumber\\
& + 4^{\alpha} \left( \frac{t^2u^2((F_{1} + F_{2})^2(k-1)^2t^2u^2 + E^2)}{D}\right)^{2\alpha}\nonumber\\
& + 64^{\alpha} \left( \frac{st^2u^2(F_{1}u + F_{2}t)(F_{1}u (s-t(k-1)) + F_{2}t(s-u(k-1)))}{D}\right)^{2\alpha}\Bigg)
\Bigg],
\end{align}
where
\begin{align*}
    D &= 2(F_{1} + F_{2})^2 (k-1)^2 t^4u^4 + (t^4 + u^4) ((F_{1} + F_{2})(k-1) tu - 2s(F_{1}u + F_{2}t))^2, \\
    E &= F_{1} u (2s -t(k-1))+F_{2}t(2s-u(k-1)),
\end{align*}
whereas for the final state~\eqref{eq:RRYMk} or~\eqref{eq:LLYMk}, it is given by
\begin{align}
M_{\alpha}\big(|\psi(k)\rangle_{RR}\big) & = \frac{1}{1-\alpha}\log \Bigg[\frac{1}{4}\Bigg(1 +2 \left(\frac{4C^2 - ((k-1)tuB)^2}{4C^{2} + ((k-1)tuB)^{2} + 2A^{2}}\right)^{2\alpha} + \left(\frac{4C^2 + ((k-1)tuB)^2 - 2A^2}{ 4C^{2} + ((k-1)tuB)^{2} + 2A^{2}} \right)^{2\alpha} \nonumber\\
& + 2^{1+2\alpha}\Bigg(\frac{A(F_{2}t(4ts^2 + 2st(t - u) + u(k-1)(u^2 - 2t^2 - 2tu))) }{4C^{2} + ((k-1)tuB)^{2} + 2A^{2}} \nonumber\\
& + \frac{A(u(F_{1}( 4us^2 + (k-1)t^3 - 2tu(s + t(k-1)) 
        + 2(s + t - kt)u^2) - F_{3}st(3k - 1)(t - u)))}{4C^{2} + ((k-1)tuB)^{2} + 2A^{2}}\Bigg)^{2\alpha} \nonumber\\
& + 2^{1+2\alpha}\Bigg(\frac{A(F_{2}t(4ts^2 + 2st(t - u) - u(k-1)(u^2 + 2t^2 + 6tu )))}{4C^{2} + ((k-1)tuB)^{2} + 2A^{2}} \nonumber\\
& + \frac{A(u(F_{1}(4us^2 - ((k-1)t^3 - 2tu(s + 3t(k-1)) + 2(s + t - kt)u^2) - F_{3}st(k + 1)(t - u)))}{4C^{2} + ((k-1)tuB)^{2} + 2A^{2}}\Bigg)^{2\alpha}\nonumber\\
& + 4^{\alpha} \left( \frac{(k-1) t u ((F_{1} + F_{2})^2 (k-1) t^3 u^3 + 2 B C)}{4C^{2} + ((k-1)tuB)^{2} + 2A^{2}}\right)^{2\alpha} + 4^{\alpha} \left( \frac{(k-1) t u ((F_{1} + F_{2})^2 (k-1) t^3 u^3 - 2 B C)}{4C^{2} + ((k-1)tuB)^{2} + 2A^{2}}\right)^{2\alpha}\Bigg)
\Bigg],
\end{align}
where
\begin{align}
    A &= (F_{1} + F_{2}) (k-1)t^2u^2 , \\
    B &= F_{1}t (t + 2u) + F_{3}s(u-t) + F_{2}u(2t + u),
    \\
    C &= u(F_{3}kst(u-t) + F_{1}u(2s^{2} + s(u-t) - (k-1)t(2t + u)))+ F_{2}t^{2}(2s^{2} + s(t - u) - (k-1)u(2u + t)).\nonumber
\end{align}

\section{Entanglement and magic-satate generation in gravitons} \label{app:details_gravitons}

In this section, we present the explicit final states generated by the graviton scattering and the corresponding analysis of entanglement and magic-state production.

We shall consider two incoming gravitons in a product state of polarizations, $|RR\rangle$, $|RL\rangle$, $|LR\rangle$ or $|LL\rangle$. .
In ordinary general relativity ($k=1$), if the initial state share the same polarization, then the interaction does not change the polarization of the gravitons and the final state remains the same product state, as can be seen by the first equation of Eq. \eqref{eq:totalGrav}, so there is no generation of entanglement in this scattering process. However, by the other two equations, when the initial product state have opposite polarizations, either $RL$ or $LR$, the interaction produces a superposition of polarization. We then restrict to the subspace of two gravitons normalizing the resulting state. In the case of an initial $|RL\rangle$ state, the above final amplitudes allow us to write the effective final state as
\begin{equation}
   |\psi\rangle_{RL \rightarrow RL + LR} =  \frac{1}{\sqrt{t^{8} + u^{8}}}\left( u^{4}|RL\rangle + t^{4}|LR\rangle\right)
   \label{eq:finalRLgrav} .
\end{equation}
In the case of an initial $|LR\rangle$ state, the result reads
\begin{equation}
   |\psi\rangle_{LR \rightarrow RL + LR} = \frac{1}{\sqrt{t^{8} + u^{8}}}\left( t^{4}|RL\rangle + u^{4}|LR\rangle\right)
   \label{eq:finalLRgrav}.
\end{equation}

We can also explore the maximal entanglement principle in this scenario in order to check whether the structure of gravitational interactions is constrained. To achieve this, we modify the balance between the 3- and 4-graviton interactions in the total amplitude applying a weight $k$ to the 4-graviton vertex.

If the initial states have opposite polarization, either $LR$ or $RL$, the corresponding normalized quantum state after the scattering interaction is 
\begin{align}
    |\psi(k)\rangle_{RL} &= \dfrac{8t^3u^3(k-1)(|RR\rangle+|LL\rangle)+s^2u^4|RL\rangle+s^2t^4|LR\rangle}{\sqrt{s^4 \left(t^8+u^8\right)+128 t^6 u^6 (k-1)^2}} \; , \label{eq:RLGravk} \\
   |\psi(k)\rangle_{LR} &= \dfrac{8t^3u^3(k-1)(|RR\rangle+|LL\rangle)+s^2t^4|RL\rangle+s^2u^4|LR\rangle}{\sqrt{s^4 \left(t^8+u^8\right)+128 t^6 u^6 (k-1)^2}} \; . \label{eq:LRGravk}
\end{align}
Fixing the COM scattering angle to $\theta = \pi/2$, their identical concurrence reads
\begin{equation}
    \Delta_{RL \rightarrow RR + RL + LR + LL} = \Delta_{LR \rightarrow RR + RL + LR + LL}  =  \left|1-\frac{2}{5+4k(k-2)}\right| .
\end{equation}
The computation shows that only the value $k=1$ leads to a final maximal entangled state with $\Delta=1$, which corresponds to the gravitational theory respecting gauge symmetry, i.e., diffeomorphism invariance. This solution is an isolated point of maximum entanglement with respect to small variations of the parameter $k$ as shown in the main article, and also suppresses the same initial polarization process, since in this case the states are given by
\begin{align}
     |\psi(k)\rangle_{RR} &=\frac{\big(s^6 +8 (k-1) t^3 u^3 \big)|RR\rangle+8 (k-1) t^3 u^3(|RL\rangle+|LR\rangle)-8 (k-1) t^2 u^2 \left(2 t^2+t u+2 u^2\right)|LL\rangle}{\sqrt{128 (k-1)^2 t^6 u^6+\left(8 (k-1) t^3 u^3+(t+u)^6\right)^2+64 (k-1)^2 t^4 u^4 \left(2 t^2+t u+2 u^2\right)^2}} \; , \label{eq:RRGravk} \\ \nonumber \\
     |\psi(k)\rangle_{LL} &= \frac{-8 (k-1) t^2 u^2 \left(2 t^2+t u+2 u^2\right)|RR\rangle+8 (k-1) t^3 u^3(|RL\rangle+|LR\rangle)+\big(s^6 +8 (k-1) t^3 u^3 \big)|LL\rangle}{\sqrt{128 (k-1)^2 t^6 u^6+\left(8 (k-1) t^3 u^3+(t+u)^6\right)^2+64 (k-1)^2 t^4 u^4 \left(2 t^2+t u+2 u^2\right)^2}} \; , \label{eq:LLGravk}
\end{align}
yielding, in the COM frame scattering angle $\theta=\pi/2$, to the identical concurrence 
\begin{equation}
    \Delta_{RR \rightarrow RR + RL + LR + LL}= \Delta_{LL \rightarrow RR + RL + LR + LL}   =  \left| \frac{(k -1)(3k+17)}{19-10k+7k^{2}}\right| .
\end{equation}

Computing the $\alpha$-order magic-state resources for the state \eqref{eq:RLGravk} or \eqref{eq:LRGravk} yields to
\begin{align*}
M_\alpha\big(|\psi(k)\rangle_{RL}\big)=&\frac{1}{1-\alpha}
\log\Bigg[
\frac{1}{4}
\Bigg(1+
\left(
\frac{
s^4 (t^8 - u^8)
}{
128 (k-1)^2 t^6 u^6 + s^4 (t^8 + u^8)
}
\right)^{2\alpha}+\, 
\left(
\frac{
s^4 (u^8 - t^8)
}{
128 (k-1)^2 t^6 u^6 + s^4 (t^8 + u^8)
}
\right)^{2\alpha}
\nonumber \\[0.6em]
+\, &
\left(
\frac{
128 (k-1)^2 t^6 u^6 - s^4 (t^8 + u^8)
}{
128 (k-1)^2 t^6 u^6 + s^4 (t^8 + u^8)
}
\right)^{2\alpha} +\, 
4^\alpha
\left(
\frac{
t^4 u^4 \left( s^4+64 (k-1)^2 t^2 u^2  \right)
}{
128 (k-1)^2 t^6 u^6 + s^4 (t^8 + u^8)
}
\right)^{2\alpha}
\nonumber \\[0.6em]
+\, &
4^\alpha
\left(
\frac{
t^4 u^4 \left( s^4 - 64 (k-1)^2 t^2 u^2 \right)
}{
128 (k-1)^2 t^6 u^6 + s^4 (t^8 + u^8)
}
\right)^{2\alpha}
+\,
256^\alpha
\left(
\frac{
(k-1) s^2 t^3 u^3 (t^4 - u^4)
}{
128 (k-1)^2 t^6 u^6 + s^4 (t^8 + u^8)
}
\right)^{2\alpha}
\nonumber \\[0.6em]
+\, &
2^{8\alpha+1}
\left(
\frac{
(k-1) s^2 t^3 u^3 (t^4 + u^4)
}{
128 (k-1)^2 t^6 u^6 + s^4 (t^8 + u^8)
}
\right)^{2\alpha} +
256^\alpha
\left(
\frac{
-(k-1) s^2 t^3 u^3 (t^4 - u^4)
}{
128 (k-1)^2 t^6 u^6 + s^4 (t^8 + u^8)
}
\right)^{2\alpha} \Bigg)\Bigg]
\end{align*}
while for the states \eqref{eq:RRGravk} or \eqref{eq:LLGravk} we obtain
\begin{align*}
M_\alpha\big(|\psi(k)\rangle_{RR}\big)&=
\frac{1}{1-\alpha} \log \Biggl[ \frac{1}{4} \Biggl( 1 + \\
    &\quad 2^{8\alpha+1} \Biggl(\frac{(k-1) t^3 u^3 \left(8 (k-1) t^3 u^3+8 (k-1) t^2 u^2 \left(2 t^2+t u+2 u^2\right)+(t+u)^6\right)}{128 (k-1)^2 t^6 u^6+\left(8 (k-1) t^3 u^3+(t+u)^6\right)^2+64 (k-1)^2 t^4 u^4 \left(2 t^2+t u+2 u^2\right)^2}\Biggr)^{2\alpha} \\
    &\quad + 2^{8\alpha+1} \Biggl(\frac{(k-1) t^3 u^3 \left(8 (k-1) t^3 u^3-8 (k-1) t^2 u^2 \left(2 t^2+t u+2 u^2\right)+(t+u)^6\right)}{128 (k-1)^2 t^6 u^6+\left(8 (k-1) t^3 u^3+(t+u)^6\right)^2+64 (k-1)^2 t^4 u^4 \left(2 t^2+t u+2 u^2\right)^2}\Biggr)^{2\alpha} \\
    &\quad + 256^\alpha \Biggl(\frac{(k-1) t^2 u^2 \left(8 (k-1) t^4 u^4+\left(2 t^2+t u+2 u^2\right) \left(8 (k-1) t^3 u^3+(t+u)^6\right)\right)}{128 (k-1)^2 t^6 u^6+\left(8 (k-1) t^3 u^3+(t+u)^6\right)^2+64 (k-1)^2 t^4 u^4 \left(2 t^2+t u+2 u^2\right)^2}\Biggr)^{2\alpha} \\
    &\quad + 256^\alpha \Biggl(\frac{(k-1) t^2 u^2 \left(8 (k-1) t^4 u^4-\left(2 t^2+t u+2 u^2\right) \left(8 (k-1) t^3 u^3+(t+u)^6\right)\right)}{128 (k-1)^2 t^6 u^6+\left(8 (k-1) t^3 u^3+(t+u)^6\right)^2+64 (k-1)^2 t^4 u^4 \left(2 t^2+t u+2 u^2\right)^2}\Biggr)^{2\alpha} \\
    &\quad + 2 \Biggl(\frac{\left(8 (k-1) t^3 u^3+(t+u)^6\right)^2-64 (k-1)^2 t^4 u^4 \left(2 t^2+t u+2 u^2\right)^2}{128 (k-1)^2 t^6 u^6+\left(8 (k-1) t^3 u^3+(t+u)^6\right)^2+64 (k-1)^2 t^4 u^4 \left(2 t^2+t u+2 u^2\right)^2}\Biggr)^{2\alpha} \\
    &\quad + \Biggl(\frac{-128 (k-1)^2 t^6 u^6+\left(8 (k-1) t^3 u^3+(t+u)^6\right)^2+64 (k-1)^2 t^4 u^4 \left(2 t^2+t u+2 u^2\right)^2}{128 (k-1)^2 t^6 u^6+\left(8 (k-1) t^3 u^3+(t+u)^6\right)^2+64 (k-1)^2 t^4 u^4 \left(2 t^2+t u+2 u^2\right)^2}\Biggr)^{2\alpha}\Biggr)\Biggr] \\
\end{align*}

\bibliography{biblio}

\begin{thebibliography}{49}%
\makeatletter
\providecommand \@ifxundefined [1]{%
 \@ifx{#1\undefined}
}%
\providecommand \@ifnum [1]{%
 \ifnum #1\expandafter \@firstoftwo
 \else \expandafter \@secondoftwo
 \fi
}%
\providecommand \@ifx [1]{%
 \ifx #1\expandafter \@firstoftwo
 \else \expandafter \@secondoftwo
 \fi
}%
\providecommand \natexlab [1]{#1}%
\providecommand \enquote  [1]{``#1''}%
\providecommand \bibnamefont  [1]{#1}%
\providecommand \bibfnamefont [1]{#1}%
\providecommand \citenamefont [1]{#1}%
\providecommand \href@noop [0]{\@secondoftwo}%
\providecommand \href [0]{\begingroup \@sanitize@url \@href}%
\providecommand \@href[1]{\@@startlink{#1}\@@href}%
\providecommand \@@href[1]{\endgroup#1\@@endlink}%
\providecommand \@sanitize@url [0]{\catcode `\\12\catcode `\$12\catcode `\&12\catcode `\#12\catcode `\^12\catcode `\_12\catcode `\%12\relax}%
\providecommand \@@startlink[1]{}%
\providecommand \@@endlink[0]{}%
\providecommand \url  [0]{\begingroup\@sanitize@url \@url }%
\providecommand \@url [1]{\endgroup\@href {#1}{\urlprefix }}%
\providecommand \urlprefix  [0]{URL }%
\providecommand \Eprint [0]{\href }%
\providecommand \doibase [0]{https://doi.org/}%
\providecommand \selectlanguage [0]{\@gobble}%
\providecommand \bibinfo  [0]{\@secondoftwo}%
\providecommand \bibfield  [0]{\@secondoftwo}%
\providecommand \translation [1]{[#1]}%
\providecommand \BibitemOpen [0]{}%
\providecommand \bibitemStop [0]{}%
\providecommand \bibitemNoStop [0]{.\EOS\space}%
\providecommand \EOS [0]{\spacefactor3000\relax}%
\providecommand \BibitemShut  [1]{\csname bibitem#1\endcsname}%
\let\auto@bib@innerbib\@empty
\bibitem [{\citenamefont {Ac\'{\i}n}\ \emph {et~al.}(2002)\citenamefont {Ac\'{\i}n}, \citenamefont {Durt}, \citenamefont {Gisin},\ and\ \citenamefont {Latorre}}]{acin2002quantum}%
  \BibitemOpen
  \bibfield  {author} {\bibinfo {author} {\bibfnamefont {A.}~\bibnamefont {Ac\'{\i}n}}, \bibinfo {author} {\bibfnamefont {T.}~\bibnamefont {Durt}}, \bibinfo {author} {\bibfnamefont {N.}~\bibnamefont {Gisin}},\ and\ \bibinfo {author} {\bibfnamefont {J.~I.}\ \bibnamefont {Latorre}},\ }\bibfield  {title} {\bibinfo {title} {Quantum nonlocality in two three-level systems},\ }\href {https://doi.org/10.1103/PhysRevA.65.052325} {\bibfield  {journal} {\bibinfo  {journal} {Phys. Rev. A}\ }\textbf {\bibinfo {volume} {65}},\ \bibinfo {pages} {052325} (\bibinfo {year} {2002})}\BibitemShut {NoStop}%
\bibitem [{\citenamefont {Ac\'{\i}n}\ \emph {et~al.}(2012)\citenamefont {Ac\'{\i}n}, \citenamefont {Massar},\ and\ \citenamefont {Pironio}}]{acin2012randomness}%
  \BibitemOpen
  \bibfield  {author} {\bibinfo {author} {\bibfnamefont {A.}~\bibnamefont {Ac\'{\i}n}}, \bibinfo {author} {\bibfnamefont {S.}~\bibnamefont {Massar}},\ and\ \bibinfo {author} {\bibfnamefont {S.}~\bibnamefont {Pironio}},\ }\bibfield  {title} {\bibinfo {title} {Randomness versus nonlocality and entanglement},\ }\href {https://doi.org/10.1103/PhysRevLett.108.100402} {\bibfield  {journal} {\bibinfo  {journal} {Phys. Rev. Lett.}\ }\textbf {\bibinfo {volume} {108}},\ \bibinfo {pages} {100402} (\bibinfo {year} {2012})}\BibitemShut {NoStop}%
\bibitem [{\citenamefont {Gottesman}(1998)}]{gottesman1998heisenberg}%
  \BibitemOpen
  \bibfield  {author} {\bibinfo {author} {\bibfnamefont {D.}~\bibnamefont {Gottesman}},\ }\bibfield  {title} {\bibinfo {title} {The heisenberg representation of quantum computers},\ }\bibfield  {journal} {\bibinfo  {journal} {arXiv:9807006 [quant-ph]}\ }\href {https://doi.org/10.48550/arXiv.quant-ph/9807006} {10.48550/arXiv.quant-ph/9807006} (\bibinfo {year} {1998})\BibitemShut {NoStop}%
\bibitem [{\citenamefont {Aaronson}\ and\ \citenamefont {Gottesman}(2004)}]{aaronson2004improved}%
  \BibitemOpen
  \bibfield  {author} {\bibinfo {author} {\bibfnamefont {S.}~\bibnamefont {Aaronson}}\ and\ \bibinfo {author} {\bibfnamefont {D.}~\bibnamefont {Gottesman}},\ }\bibfield  {title} {\bibinfo {title} {Improved simulation of stabilizer circuits},\ }\href {https://doi.org/10.1103/PhysRevA.70.052328} {\bibfield  {journal} {\bibinfo  {journal} {Phys. Rev. A}\ }\textbf {\bibinfo {volume} {70}},\ \bibinfo {pages} {052328} (\bibinfo {year} {2004})}\BibitemShut {NoStop}%
\bibitem [{\citenamefont {Iannotti}\ \emph {et~al.}(2025)\citenamefont {Iannotti}, \citenamefont {Esposito}, \citenamefont {Venuti},\ and\ \citenamefont {Hamma}}]{iannotti2025entanglement}%
  \BibitemOpen
  \bibfield  {author} {\bibinfo {author} {\bibfnamefont {D.}~\bibnamefont {Iannotti}}, \bibinfo {author} {\bibfnamefont {G.}~\bibnamefont {Esposito}}, \bibinfo {author} {\bibfnamefont {L.~C.}\ \bibnamefont {Venuti}},\ and\ \bibinfo {author} {\bibfnamefont {A.}~\bibnamefont {Hamma}},\ }\bibfield  {title} {\bibinfo {title} {Entanglement and stabilizer entropies of random bipartite pure quantum states},\ }\href {https://doi.org/10.22331/q-2025-07-21-1797} {\bibfield  {journal} {\bibinfo  {journal} {Quantum 9, 1797 (2025).}\ }\textbf {\bibinfo {volume} {9}},\ \bibinfo {pages} {1797} (\bibinfo {year} {2025})}\BibitemShut {NoStop}%
\bibitem [{\citenamefont {Cervera-Lierta}\ \emph {et~al.}(2017)\citenamefont {Cervera-Lierta}, \citenamefont {Latorre}, \citenamefont {Rojo},\ and\ \citenamefont {Rottoli}}]{CerveraLierta2017maximal}%
  \BibitemOpen
  \bibfield  {author} {\bibinfo {author} {\bibfnamefont {A.}~\bibnamefont {Cervera-Lierta}}, \bibinfo {author} {\bibfnamefont {J.~I.}\ \bibnamefont {Latorre}}, \bibinfo {author} {\bibfnamefont {J.}~\bibnamefont {Rojo}},\ and\ \bibinfo {author} {\bibfnamefont {L.}~\bibnamefont {Rottoli}},\ }\bibfield  {title} {\bibinfo {title} {{Maximal Entanglement in High Energy Physics}},\ }\href {https://doi.org/10.21468/SciPostPhys.3.5.036} {\bibfield  {journal} {\bibinfo  {journal} {SciPost Phys.}\ }\textbf {\bibinfo {volume} {3}},\ \bibinfo {pages} {036} (\bibinfo {year} {2017})}\BibitemShut {NoStop}%
\bibitem [{\citenamefont {Morales}(2024)}]{Morales2024tripartite}%
  \BibitemOpen
  \bibfield  {author} {\bibinfo {author} {\bibfnamefont {R.~A.}\ \bibnamefont {Morales}},\ }\bibfield  {title} {\bibinfo {title} {{{Tripartite entanglement and Bell non-locality in loop-induced Higgs boson decays}}},\ }\href {https://doi.org/10.1140/epjc/s10052-024-12921-4} {\bibfield  {journal} {\bibinfo  {journal} {Eur. Phys. J. C}\ }\textbf {\bibinfo {volume} {84}},\ \bibinfo {pages} {581} (\bibinfo {year} {2024})},\ \Eprint {https://arxiv.org/abs/2403.18023} {arXiv:2403.18023 [hep-ph]} \BibitemShut {NoStop}%
\bibitem [{\citenamefont {N{\'u}{\~n}ez}\ \emph {et~al.}(2026)\citenamefont {N{\'u}{\~n}ez}, \citenamefont {Cervera-Lierta},\ and\ \citenamefont {Latorre}}]{nunez2025universality}%
  \BibitemOpen
  \bibfield  {author} {\bibinfo {author} {\bibfnamefont {C.}~\bibnamefont {N{\'u}{\~n}ez}}, \bibinfo {author} {\bibfnamefont {A.}~\bibnamefont {Cervera-Lierta}},\ and\ \bibinfo {author} {\bibfnamefont {J.~I.}\ \bibnamefont {Latorre}},\ }\bibfield  {title} {\bibinfo {title} {Universality of entanglement in gluon dynamics},\ }\href {https://doi.org/10.21468/SciPostPhys.20.2.063} {\bibfield  {journal} {\bibinfo  {journal} {SciPost Phys.}\ }\textbf {\bibinfo {volume} {20}},\ \bibinfo {pages} {063} (\bibinfo {year} {2026})}\BibitemShut {NoStop}%
\bibitem [{\citenamefont {{ATLAS Collaboration}}(2024)}]{ATLASCollaboration2024observation}%
  \BibitemOpen
  \bibfield  {author} {\bibinfo {author} {\bibnamefont {{ATLAS Collaboration}}},\ }\bibfield  {title} {\bibinfo {title} {{Observation of quantum entanglement with top quarks at the ATLAS detector}},\ }\href {https://doi.org/10.1038/s41586-024-07824-z} {\bibfield  {journal} {\bibinfo  {journal} {Nature}\ }\textbf {\bibinfo {volume} {633}},\ \bibinfo {pages} {542–547} (\bibinfo {year} {2024})}\BibitemShut {NoStop}%
\bibitem [{\citenamefont {{CMS Collaboration}}(2024{\natexlab{a}})}]{CMSCollaboration2024observation}%
  \BibitemOpen
  \bibfield  {author} {\bibinfo {author} {\bibnamefont {{CMS Collaboration}}},\ }\bibfield  {title} {\bibinfo {title} {{Observation of quantum entanglement in top quark pair production in proton–proton collisions at $\sqrt{s}$ = 13 TeV}},\ }\href {https://doi.org/10.1088/1361-6633/ad7e4d} {\bibfield  {journal} {\bibinfo  {journal} {Reports on Progress in Physics}\ }\textbf {\bibinfo {volume} {87}},\ \bibinfo {pages} {117801} (\bibinfo {year} {2024}{\natexlab{a}})}\BibitemShut {NoStop}%
\bibitem [{\citenamefont {{CMS Collaboration}}(2024{\natexlab{b}})}]{CMSCollaboration2024measurements}%
  \BibitemOpen
  \bibfield  {author} {\bibinfo {author} {\bibnamefont {{CMS Collaboration}}},\ }\bibfield  {title} {\bibinfo {title} {{Measurements of polarization and spin correlation and observation of entanglement in top quark pairs using $\mathrm{lepton}+\text{jets}$ events from proton-proton collisions at $\sqrt{s}=13\text{ }\text{ }\mathrm{TeV}$}},\ }\href {https://doi.org/10.1103/PhysRevD.110.112016} {\bibfield  {journal} {\bibinfo  {journal} {Phys. Rev. D}\ }\textbf {\bibinfo {volume} {110}},\ \bibinfo {pages} {112016} (\bibinfo {year} {2024}{\natexlab{b}})}\BibitemShut {NoStop}%
\bibitem [{\citenamefont {collaboration}(2025)}]{yazgan2025measurements}%
  \BibitemOpen
  \bibfield  {author} {\bibinfo {author} {\bibfnamefont {C.}~\bibnamefont {collaboration}},\ }\bibfield  {title} {\bibinfo {title} {Measurements of top quark properties in cms: {$t\bar{t}$} spin density matrix, quantum entanglement and quantum magic},\ }\bibfield  {journal} {\bibinfo  {journal} {arXiv:2510.13743 [hep-ex]}\ }\href {https://doi.org/10.48550/arXiv.2510.13743} {10.48550/arXiv.2510.13743} (\bibinfo {year} {2025})\BibitemShut {NoStop}%
\bibitem [{\citenamefont {Fabbrichesi}\ \emph {et~al.}(2024{\natexlab{a}})\citenamefont {Fabbrichesi}, \citenamefont {Floreanini}, \citenamefont {Gabrielli},\ and\ \citenamefont {Marzola}}]{Fabbrichesi2024charmonium}%
  \BibitemOpen
  \bibfield  {author} {\bibinfo {author} {\bibfnamefont {M.}~\bibnamefont {Fabbrichesi}}, \bibinfo {author} {\bibfnamefont {R.}~\bibnamefont {Floreanini}}, \bibinfo {author} {\bibfnamefont {E.}~\bibnamefont {Gabrielli}},\ and\ \bibinfo {author} {\bibfnamefont {L.}~\bibnamefont {Marzola}},\ }\bibfield  {title} {\bibinfo {title} {{Bell inequality is violated in charmonium decays}},\ }\href {https://doi.org/10.1103/PhysRevD.110.053008} {\bibfield  {journal} {\bibinfo  {journal} {Phys. Rev. D}\ }\textbf {\bibinfo {volume} {110}},\ \bibinfo {pages} {053008} (\bibinfo {year} {2024}{\natexlab{a}})}\BibitemShut {NoStop}%
\bibitem [{\citenamefont {Fabbrichesi}\ \emph {et~al.}(2024{\natexlab{b}})\citenamefont {Fabbrichesi}, \citenamefont {Floreanini}, \citenamefont {Gabrielli},\ and\ \citenamefont {Marzola}}]{Fabbrichesi2024bell}%
  \BibitemOpen
  \bibfield  {author} {\bibinfo {author} {\bibfnamefont {M.}~\bibnamefont {Fabbrichesi}}, \bibinfo {author} {\bibfnamefont {R.}~\bibnamefont {Floreanini}}, \bibinfo {author} {\bibfnamefont {E.}~\bibnamefont {Gabrielli}},\ and\ \bibinfo {author} {\bibfnamefont {L.}~\bibnamefont {Marzola}},\ }\bibfield  {title} {\bibinfo {title} {{Bell inequality is violated in ${B}^{0}\ensuremath{\rightarrow}J/\ensuremath{\psi}{K}^{*}(892{)}^{0}$ decays}},\ }\href {https://doi.org/10.1103/PhysRevD.109.L031104} {\bibfield  {journal} {\bibinfo  {journal} {Phys. Rev. D}\ }\textbf {\bibinfo {volume} {109}},\ \bibinfo {pages} {L031104} (\bibinfo {year} {2024}{\natexlab{b}})}\BibitemShut {NoStop}%
\bibitem [{\citenamefont {Gabrielli}\ and\ \citenamefont {Marzola}(2024)}]{Gabrielli2024entanglement}%
  \BibitemOpen
  \bibfield  {author} {\bibinfo {author} {\bibfnamefont {E.}~\bibnamefont {Gabrielli}}\ and\ \bibinfo {author} {\bibfnamefont {L.}~\bibnamefont {Marzola}},\ }\bibfield  {title} {\bibinfo {title} {{Entanglement and Bell Inequality Violation in $B \ensuremath{\rightarrow}\ensuremath{\Phi}\ensuremath{\Phi}$ Decays}},\ }\href {https://doi.org/10.3390/sym16081036} {\bibfield  {journal} {\bibinfo  {journal} {Symmetry}\ }\textbf {\bibinfo {volume} {16}},\ \bibinfo {pages} {1036} (\bibinfo {year} {2024})}\BibitemShut {NoStop}%
\bibitem [{\citenamefont {Beane}\ \emph {et~al.}(2019)\citenamefont {Beane}, \citenamefont {Kaplan}, \citenamefont {Klco},\ and\ \citenamefont {Savage}}]{Beane2019entanglement}%
  \BibitemOpen
  \bibfield  {author} {\bibinfo {author} {\bibfnamefont {S.~R.}\ \bibnamefont {Beane}}, \bibinfo {author} {\bibfnamefont {D.~B.}\ \bibnamefont {Kaplan}}, \bibinfo {author} {\bibfnamefont {N.}~\bibnamefont {Klco}},\ and\ \bibinfo {author} {\bibfnamefont {M.~J.}\ \bibnamefont {Savage}},\ }\bibfield  {title} {\bibinfo {title} {{Entanglement Suppression and Emergent Symmetries of Strong Interactions}},\ }\href {https://doi.org/10.1103/PhysRevLett.122.102001} {\bibfield  {journal} {\bibinfo  {journal} {Phys. Rev. Lett.}\ }\textbf {\bibinfo {volume} {122}},\ \bibinfo {pages} {102001} (\bibinfo {year} {2019})}\BibitemShut {NoStop}%
\bibitem [{\citenamefont {Low}\ and\ \citenamefont {Mehen}(2021)}]{Low2021symmetry}%
  \BibitemOpen
  \bibfield  {author} {\bibinfo {author} {\bibfnamefont {I.}~\bibnamefont {Low}}\ and\ \bibinfo {author} {\bibfnamefont {T.}~\bibnamefont {Mehen}},\ }\bibfield  {title} {\bibinfo {title} {{Symmetry from entanglement suppression}},\ }\href {https://doi.org/10.1103/PhysRevD.104.074014} {\bibfield  {journal} {\bibinfo  {journal} {Phys. Rev. D}\ }\textbf {\bibinfo {volume} {104}},\ \bibinfo {pages} {074014} (\bibinfo {year} {2021})}\BibitemShut {NoStop}%
\bibitem [{\citenamefont {Liu}\ \emph {et~al.}(2023)\citenamefont {Liu}, \citenamefont {Low},\ and\ \citenamefont {Mehen}}]{Liu2023minimal}%
  \BibitemOpen
  \bibfield  {author} {\bibinfo {author} {\bibfnamefont {Q.}~\bibnamefont {Liu}}, \bibinfo {author} {\bibfnamefont {I.}~\bibnamefont {Low}},\ and\ \bibinfo {author} {\bibfnamefont {T.}~\bibnamefont {Mehen}},\ }\bibfield  {title} {\bibinfo {title} {{Minimal entanglement and emergent symmetries in low-energy QCD}},\ }\href {https://doi.org/10.1103/PhysRevC.107.025204} {\bibfield  {journal} {\bibinfo  {journal} {Phys. Rev. C}\ }\textbf {\bibinfo {volume} {107}},\ \bibinfo {pages} {025204} (\bibinfo {year} {2023})}\BibitemShut {NoStop}%
\bibitem [{\citenamefont {Carena}\ \emph {et~al.}(2025)\citenamefont {Carena}, \citenamefont {Coloretti}, \citenamefont {Liu}, \citenamefont {Littmann}, \citenamefont {Low},\ and\ \citenamefont {Wagner}}]{carena2025entanglement}%
  \BibitemOpen
  \bibfield  {author} {\bibinfo {author} {\bibfnamefont {M.}~\bibnamefont {Carena}}, \bibinfo {author} {\bibfnamefont {G.}~\bibnamefont {Coloretti}}, \bibinfo {author} {\bibfnamefont {W.}~\bibnamefont {Liu}}, \bibinfo {author} {\bibfnamefont {M.}~\bibnamefont {Littmann}}, \bibinfo {author} {\bibfnamefont {I.}~\bibnamefont {Low}},\ and\ \bibinfo {author} {\bibfnamefont {C.~E.}\ \bibnamefont {Wagner}},\ }\bibfield  {title} {\bibinfo {title} {Entanglement maximization and mirror symmetry in two-higgs-doublet models},\ }\bibfield  {journal} {\bibinfo  {journal} {arXiv:2505.00873 [hep-ph]}\ }\href {https://doi.org/10.48550/arXiv.2505.00873} {10.48550/arXiv.2505.00873} (\bibinfo {year} {2025})\BibitemShut {NoStop}%
\bibitem [{\citenamefont {Thaler}\ and\ \citenamefont {Trifinopoulos}(2025)}]{PhysRevD.111.056021}%
  \BibitemOpen
  \bibfield  {author} {\bibinfo {author} {\bibfnamefont {J.}~\bibnamefont {Thaler}}\ and\ \bibinfo {author} {\bibfnamefont {S.}~\bibnamefont {Trifinopoulos}},\ }\bibfield  {title} {\bibinfo {title} {Flavor patterns of fundamental particles from quantum entanglement?},\ }\href {https://doi.org/10.1103/PhysRevD.111.056021} {\bibfield  {journal} {\bibinfo  {journal} {Phys. Rev. D}\ }\textbf {\bibinfo {volume} {111}},\ \bibinfo {pages} {056021} (\bibinfo {year} {2025})}\BibitemShut {NoStop}%
\bibitem [{\citenamefont {McGinnis}(2025)}]{mcginnis2025crossing}%
  \BibitemOpen
  \bibfield  {author} {\bibinfo {author} {\bibfnamefont {N.}~\bibnamefont {McGinnis}},\ }\bibfield  {title} {\bibinfo {title} {Crossing symmetry and entanglement},\ }\bibfield  {journal} {\bibinfo  {journal} {arXiv:2511.10559 [hep-th]}\ }\href {https://doi.org/10.48550/arXiv.2511.10559} {10.48550/arXiv.2511.10559} (\bibinfo {year} {2025})\BibitemShut {NoStop}%
\bibitem [{\citenamefont {Liu}\ \emph {et~al.}(2025{\natexlab{a}})\citenamefont {Liu}, \citenamefont {Low},\ and\ \citenamefont {Yin}}]{liu2025qed}%
  \BibitemOpen
  \bibfield  {author} {\bibinfo {author} {\bibfnamefont {Q.}~\bibnamefont {Liu}}, \bibinfo {author} {\bibfnamefont {I.}~\bibnamefont {Low}},\ and\ \bibinfo {author} {\bibfnamefont {Z.}~\bibnamefont {Yin}},\ }\bibfield  {title} {\bibinfo {title} {Quantum magic in quantum electrodynamics},\ }\bibfield  {journal} {\bibinfo  {journal} {arXiv:2503.03098 [hep-th]}\ }\href {https://doi.org/10.48550/arXiv.2503.03098} {10.48550/arXiv.2503.03098} (\bibinfo {year} {2025}{\natexlab{a}})\BibitemShut {NoStop}%
\bibitem [{\citenamefont {White}\ and\ \citenamefont {White}(2024)}]{white2024magic}%
  \BibitemOpen
  \bibfield  {author} {\bibinfo {author} {\bibfnamefont {C.~D.}\ \bibnamefont {White}}\ and\ \bibinfo {author} {\bibfnamefont {M.~J.}\ \bibnamefont {White}},\ }\bibfield  {title} {\bibinfo {title} {Magic states of top quarks},\ }\href {https://doi.org/10.1103/PhysRevD.110.116016} {\bibfield  {journal} {\bibinfo  {journal} {Phys. Rev. D}\ }\textbf {\bibinfo {volume} {110}},\ \bibinfo {pages} {116016} (\bibinfo {year} {2024})}\BibitemShut {NoStop}%
\bibitem [{\citenamefont {Gargalionis}\ \emph {et~al.}(2025)\citenamefont {Gargalionis}, \citenamefont {Moynihan}, \citenamefont {Trifinopoulos}, \citenamefont {Wallace}, \citenamefont {White},\ and\ \citenamefont {White}}]{gargalionis2025spin}%
  \BibitemOpen
  \bibfield  {author} {\bibinfo {author} {\bibfnamefont {J.}~\bibnamefont {Gargalionis}}, \bibinfo {author} {\bibfnamefont {N.}~\bibnamefont {Moynihan}}, \bibinfo {author} {\bibfnamefont {S.}~\bibnamefont {Trifinopoulos}}, \bibinfo {author} {\bibfnamefont {E.~N.}\ \bibnamefont {Wallace}}, \bibinfo {author} {\bibfnamefont {C.~D.}\ \bibnamefont {White}},\ and\ \bibinfo {author} {\bibfnamefont {M.~J.}\ \bibnamefont {White}},\ }\bibfield  {title} {\bibinfo {title} {Spin versus magic: Lessons from gluon and graviton scattering},\ }\bibfield  {journal} {\bibinfo  {journal} {arXiv:2508.14967 [hep-th]}\ }\href {https://doi.org/10.48550/arXiv.2508.14967} {10.48550/arXiv.2508.14967} (\bibinfo {year} {2025})\BibitemShut {NoStop}%
\bibitem [{\citenamefont {Tarabunga}\ \emph {et~al.}(2023)\citenamefont {Tarabunga}, \citenamefont {Tirrito}, \citenamefont {Chanda},\ and\ \citenamefont {Dalmonte}}]{PRXQuantum.4.040317}%
  \BibitemOpen
  \bibfield  {author} {\bibinfo {author} {\bibfnamefont {P.~S.}\ \bibnamefont {Tarabunga}}, \bibinfo {author} {\bibfnamefont {E.}~\bibnamefont {Tirrito}}, \bibinfo {author} {\bibfnamefont {T.}~\bibnamefont {Chanda}},\ and\ \bibinfo {author} {\bibfnamefont {M.}~\bibnamefont {Dalmonte}},\ }\bibfield  {title} {\bibinfo {title} {Many-body magic via pauli-markov chains---from criticality to gauge theories},\ }\href {https://doi.org/10.1103/PRXQuantum.4.040317} {\bibfield  {journal} {\bibinfo  {journal} {PRX Quantum}\ }\textbf {\bibinfo {volume} {4}},\ \bibinfo {pages} {040317} (\bibinfo {year} {2023})}\BibitemShut {NoStop}%
\bibitem [{\citenamefont {Liu}\ \emph {et~al.}(2025{\natexlab{b}})\citenamefont {Liu}, \citenamefont {Low},\ and\ \citenamefont {Yin}}]{liu2025quantum}%
  \BibitemOpen
  \bibfield  {author} {\bibinfo {author} {\bibfnamefont {Q.}~\bibnamefont {Liu}}, \bibinfo {author} {\bibfnamefont {I.}~\bibnamefont {Low}},\ and\ \bibinfo {author} {\bibfnamefont {Z.}~\bibnamefont {Yin}},\ }\bibfield  {title} {\bibinfo {title} {A quantum computational determination of the weak mixing angle in the standard model},\ }\bibfield  {journal} {\bibinfo  {journal} {arXiv:2509.18251 [hep-th]}\ }\href {https://doi.org/10.48550/arXiv.2509.18251} {10.48550/arXiv.2509.18251} (\bibinfo {year} {2025}{\natexlab{b}})\BibitemShut {NoStop}%
\bibitem [{\citenamefont {Aoude}\ \emph {et~al.}(2022)\citenamefont {Aoude}, \citenamefont {Madge}, \citenamefont {Maltoni},\ and\ \citenamefont {Mantani}}]{Aoude2022quantum}%
  \BibitemOpen
  \bibfield  {author} {\bibinfo {author} {\bibfnamefont {R.}~\bibnamefont {Aoude}}, \bibinfo {author} {\bibfnamefont {E.}~\bibnamefont {Madge}}, \bibinfo {author} {\bibfnamefont {F.}~\bibnamefont {Maltoni}},\ and\ \bibinfo {author} {\bibfnamefont {L.}~\bibnamefont {Mantani}},\ }\bibfield  {title} {\bibinfo {title} {{Quantum SMEFT tomography: Top quark pair production at the LHC}},\ }\href {https://doi.org/10.1103/PhysRevD.106.055007} {\bibfield  {journal} {\bibinfo  {journal} {Phys. Rev. D}\ }\textbf {\bibinfo {volume} {106}},\ \bibinfo {pages} {055007} (\bibinfo {year} {2022})}\BibitemShut {NoStop}%
\bibitem [{\citenamefont {Severi}\ and\ \citenamefont {Vryonidou}(2023)}]{Severi2023quantum}%
  \BibitemOpen
  \bibfield  {author} {\bibinfo {author} {\bibfnamefont {C.}~\bibnamefont {Severi}}\ and\ \bibinfo {author} {\bibfnamefont {E.}~\bibnamefont {Vryonidou}},\ }\bibfield  {title} {\bibinfo {title} {{Quantum entanglement and top spin correlations in SMEFT at higher orders}},\ }\href {https://doi.org/10.1007/jhep01(2023)148} {\bibfield  {journal} {\bibinfo  {journal} {Journal of High Energy Physics}\ }\textbf {\bibinfo {volume} {2023}},\ \bibinfo {pages} {148} (\bibinfo {year} {2023})}\BibitemShut {NoStop}%
\bibitem [{\citenamefont {Aoude}\ \emph {et~al.}(2023)\citenamefont {Aoude}, \citenamefont {Madge}, \citenamefont {Maltoni},\ and\ \citenamefont {Mantani}}]{Aoude2023probing}%
  \BibitemOpen
  \bibfield  {author} {\bibinfo {author} {\bibfnamefont {R.}~\bibnamefont {Aoude}}, \bibinfo {author} {\bibfnamefont {E.}~\bibnamefont {Madge}}, \bibinfo {author} {\bibfnamefont {F.}~\bibnamefont {Maltoni}},\ and\ \bibinfo {author} {\bibfnamefont {L.}~\bibnamefont {Mantani}},\ }\bibfield  {title} {\bibinfo {title} {{Probing new physics through entanglement in diboson production}},\ }\href {https://doi.org/10.1007/jhep12(2023)017} {\bibfield  {journal} {\bibinfo  {journal} {Journal of High Energy Physics}\ }\textbf {\bibinfo {volume} {2023}},\ \bibinfo {pages} {17} (\bibinfo {year} {2023})}\BibitemShut {NoStop}%
\bibitem [{\citenamefont {Bernal}\ \emph {et~al.}(2023)\citenamefont {Bernal}, \citenamefont {Caban},\ and\ \citenamefont {Rembieliński}}]{Bernal2023entanglement}%
  \BibitemOpen
  \bibfield  {author} {\bibinfo {author} {\bibfnamefont {A.}~\bibnamefont {Bernal}}, \bibinfo {author} {\bibfnamefont {P.}~\bibnamefont {Caban}},\ and\ \bibinfo {author} {\bibfnamefont {J.}~\bibnamefont {Rembieliński}},\ }\bibfield  {title} {\bibinfo {title} {{Entanglement and Bell inequalities violation in $H\ensuremath{\rightarrow} ZZ$ with anomalous coupling}},\ }\href {https://doi.org/10.1140/epjc/s10052-023-12216-0} {\bibfield  {journal} {\bibinfo  {journal} {The European Physical Journal C}\ }\textbf {\bibinfo {volume} {83}},\ \bibinfo {pages} {1050} (\bibinfo {year} {2023})}\BibitemShut {NoStop}%
\bibitem [{\citenamefont {Fabbrichesi}\ \emph {et~al.}(2023)\citenamefont {Fabbrichesi}, \citenamefont {Floreanini}, \citenamefont {Gabrielli},\ and\ \citenamefont {Marzola}}]{Fabbrichesi2023stringent}%
  \BibitemOpen
  \bibfield  {author} {\bibinfo {author} {\bibfnamefont {M.}~\bibnamefont {Fabbrichesi}}, \bibinfo {author} {\bibfnamefont {R.}~\bibnamefont {Floreanini}}, \bibinfo {author} {\bibfnamefont {E.}~\bibnamefont {Gabrielli}},\ and\ \bibinfo {author} {\bibfnamefont {L.}~\bibnamefont {Marzola}},\ }\bibfield  {title} {\bibinfo {title} {{Stringent bounds on HWW and HZZ anomalous couplings with quantum tomography at the LHC}},\ }\href {https://doi.org/10.1007/jhep09(2023)195} {\bibfield  {journal} {\bibinfo  {journal} {Journal of High Energy Physics}\ }\textbf {\bibinfo {volume} {2023}},\ \bibinfo {pages} {195} (\bibinfo {year} {2023})}\BibitemShut {NoStop}%
\bibitem [{\citenamefont {Maltoni}\ \emph {et~al.}(2024{\natexlab{a}})\citenamefont {Maltoni}, \citenamefont {Severi}, \citenamefont {Tentori},\ and\ \citenamefont {Vryonidou}}]{Maltoni2024quantum}%
  \BibitemOpen
  \bibfield  {author} {\bibinfo {author} {\bibfnamefont {F.}~\bibnamefont {Maltoni}}, \bibinfo {author} {\bibfnamefont {C.}~\bibnamefont {Severi}}, \bibinfo {author} {\bibfnamefont {S.}~\bibnamefont {Tentori}},\ and\ \bibinfo {author} {\bibfnamefont {E.}~\bibnamefont {Vryonidou}},\ }\bibfield  {title} {\bibinfo {title} {{Quantum detection of new physics in top-quark pair production at the LHC}},\ }\href {https://doi.org/10.1007/jhep03(2024)099} {\bibfield  {journal} {\bibinfo  {journal} {Journal of High Energy Physics}\ }\textbf {\bibinfo {volume} {2024}},\ \bibinfo {pages} {99} (\bibinfo {year} {2024}{\natexlab{a}})}\BibitemShut {NoStop}%
\bibitem [{\citenamefont {Maltoni}\ \emph {et~al.}(2024{\natexlab{b}})\citenamefont {Maltoni}, \citenamefont {Severi}, \citenamefont {Tentori},\ and\ \citenamefont {Vryonidou}}]{Maltoni2024tops}%
  \BibitemOpen
  \bibfield  {author} {\bibinfo {author} {\bibfnamefont {F.}~\bibnamefont {Maltoni}}, \bibinfo {author} {\bibfnamefont {C.}~\bibnamefont {Severi}}, \bibinfo {author} {\bibfnamefont {S.}~\bibnamefont {Tentori}},\ and\ \bibinfo {author} {\bibfnamefont {E.}~\bibnamefont {Vryonidou}},\ }\bibfield  {title} {\bibinfo {title} {{Quantum tops at circular lepton colliders}},\ }\href {https://doi.org/10.1007/jhep09(2024)001} {\bibfield  {journal} {\bibinfo  {journal} {Journal of High Energy Physics}\ }\textbf {\bibinfo {volume} {2024}},\ \bibinfo {pages} {1} (\bibinfo {year} {2024}{\natexlab{b}})}\BibitemShut {NoStop}%
\bibitem [{\citenamefont {Carena}\ \emph {et~al.}(2023)\citenamefont {Carena}, \citenamefont {Low}, \citenamefont {Wagner},\ and\ \citenamefont {Xiao}}]{Carena2023entanglement}%
  \BibitemOpen
  \bibfield  {author} {\bibinfo {author} {\bibfnamefont {M.}~\bibnamefont {Carena}}, \bibinfo {author} {\bibfnamefont {I.}~\bibnamefont {Low}}, \bibinfo {author} {\bibfnamefont {C.~E.~M.}\ \bibnamefont {Wagner}},\ and\ \bibinfo {author} {\bibfnamefont {M.-L.}\ \bibnamefont {Xiao}},\ }\bibfield  {title} {\bibinfo {title} {{Entanglement Suppression, Enhanced Symmetry and a Standard-Model-like Higgs Boson}},\ }\bibfield  {journal} {\bibinfo  {journal} {arXiv:2307.08112 [hep-ph]}\ }\href {https://doi.org/10.48550/arXiv.2307.08112} {10.48550/arXiv.2307.08112} (\bibinfo {year} {2023})\BibitemShut {NoStop}%
\bibitem [{\citenamefont {Kowalska}\ and\ \citenamefont {Sessolo}(2024)}]{Kowalska2024entanglement}%
  \BibitemOpen
  \bibfield  {author} {\bibinfo {author} {\bibfnamefont {K.}~\bibnamefont {Kowalska}}\ and\ \bibinfo {author} {\bibfnamefont {E.~M.}\ \bibnamefont {Sessolo}},\ }\bibfield  {title} {\bibinfo {title} {{Entanglement in flavored scalar scattering}},\ }\href {https://doi.org/10.1007/jhep07(2024)156} {\bibfield  {journal} {\bibinfo  {journal} {Journal of High Energy Physics}\ }\textbf {\bibinfo {volume} {2024}},\ \bibinfo {pages} {156} (\bibinfo {year} {2024})}\BibitemShut {NoStop}%
\bibitem [{\citenamefont {Horodecki}\ \emph {et~al.}(1995)\citenamefont {Horodecki}, \citenamefont {Horodecki},\ and\ \citenamefont {Horodecki}}]{horodecki1995violating}%
  \BibitemOpen
  \bibfield  {author} {\bibinfo {author} {\bibfnamefont {R.}~\bibnamefont {Horodecki}}, \bibinfo {author} {\bibfnamefont {P.}~\bibnamefont {Horodecki}},\ and\ \bibinfo {author} {\bibfnamefont {M.}~\bibnamefont {Horodecki}},\ }\bibfield  {title} {\bibinfo {title} {Violating bell inequality by mixed spin-12 states: necessary and sufficient condition},\ }\href {https://doi.org/10.1016/0375-9601(95)00214-N} {\bibfield  {journal} {\bibinfo  {journal} {Physics Letters A}\ }\textbf {\bibinfo {volume} {200}},\ \bibinfo {pages} {340} (\bibinfo {year} {1995})}\BibitemShut {NoStop}%
\bibitem [{\citenamefont {Nielsen}(1999)}]{Nielsen1991conditions}%
  \BibitemOpen
  \bibfield  {author} {\bibinfo {author} {\bibfnamefont {M.~A.}\ \bibnamefont {Nielsen}},\ }\bibfield  {title} {\bibinfo {title} {Conditions for a class of entanglement transformations},\ }\href {https://doi.org/10.1103/PhysRevLett.83.436} {\bibfield  {journal} {\bibinfo  {journal} {Phys. Rev. Lett.}\ }\textbf {\bibinfo {volume} {83}},\ \bibinfo {pages} {436} (\bibinfo {year} {1999})}\BibitemShut {NoStop}%
\bibitem [{\citenamefont {Helwig}\ and\ \citenamefont {Cui}(2013)}]{helwig2013absolutely}%
  \BibitemOpen
  \bibfield  {author} {\bibinfo {author} {\bibfnamefont {W.}~\bibnamefont {Helwig}}\ and\ \bibinfo {author} {\bibfnamefont {W.}~\bibnamefont {Cui}},\ }\bibfield  {title} {\bibinfo {title} {Absolutely maximally entangled states: existence and applications},\ }\bibfield  {journal} {\bibinfo  {journal} {arXiv:1306.2536 [quant-ph]}\ }\href {https://doi.org/10.48550/arXiv.1306.2536} {10.48550/arXiv.1306.2536} (\bibinfo {year} {2013})\BibitemShut {NoStop}%
\bibitem [{\citenamefont {Cervera-Lierta}\ \emph {et~al.}(2019)\citenamefont {Cervera-Lierta}, \citenamefont {Latorre},\ and\ \citenamefont {Goyeneche}}]{Cervera2019quantum}%
  \BibitemOpen
  \bibfield  {author} {\bibinfo {author} {\bibfnamefont {A.}~\bibnamefont {Cervera-Lierta}}, \bibinfo {author} {\bibfnamefont {J.~I.}\ \bibnamefont {Latorre}},\ and\ \bibinfo {author} {\bibfnamefont {D.}~\bibnamefont {Goyeneche}},\ }\bibfield  {title} {\bibinfo {title} {Quantum circuits for maximally entangled states},\ }\href {https://doi.org/10.1103/PhysRevA.100.022342} {\bibfield  {journal} {\bibinfo  {journal} {Phys. Rev. A}\ }\textbf {\bibinfo {volume} {100}},\ \bibinfo {pages} {022342} (\bibinfo {year} {2019})}\BibitemShut {NoStop}%
\bibitem [{\citenamefont {Hein}\ \emph {et~al.}(2004)\citenamefont {Hein}, \citenamefont {Eisert},\ and\ \citenamefont {Briegel}}]{hein2004multiparty}%
  \BibitemOpen
  \bibfield  {author} {\bibinfo {author} {\bibfnamefont {M.}~\bibnamefont {Hein}}, \bibinfo {author} {\bibfnamefont {J.}~\bibnamefont {Eisert}},\ and\ \bibinfo {author} {\bibfnamefont {H.~J.}\ \bibnamefont {Briegel}},\ }\bibfield  {title} {\bibinfo {title} {Multiparty entanglement in graph states},\ }\href {https://doi.org/10.1103/PhysRevA.69.062311} {\bibfield  {journal} {\bibinfo  {journal} {Phys. Rev. A}\ }\textbf {\bibinfo {volume} {69}},\ \bibinfo {pages} {062311} (\bibinfo {year} {2004})}\BibitemShut {NoStop}%
\bibitem [{\citenamefont {Liu}\ and\ \citenamefont {Winter}(2022)}]{liu2022many}%
  \BibitemOpen
  \bibfield  {author} {\bibinfo {author} {\bibfnamefont {Z.-W.}\ \bibnamefont {Liu}}\ and\ \bibinfo {author} {\bibfnamefont {A.}~\bibnamefont {Winter}},\ }\bibfield  {title} {\bibinfo {title} {Many-body quantum magic},\ }\href {https://doi.org/10.1103/PRXQuantum.3.020333} {\bibfield  {journal} {\bibinfo  {journal} {PRX Quantum}\ }\textbf {\bibinfo {volume} {3}},\ \bibinfo {pages} {020333} (\bibinfo {year} {2022})}\BibitemShut {NoStop}%
\bibitem [{\citenamefont {Leone}\ \emph {et~al.}(2022)\citenamefont {Leone}, \citenamefont {Oliviero},\ and\ \citenamefont {Hamma}}]{leone2022stabilizer}%
  \BibitemOpen
  \bibfield  {author} {\bibinfo {author} {\bibfnamefont {L.}~\bibnamefont {Leone}}, \bibinfo {author} {\bibfnamefont {S.~F.~E.}\ \bibnamefont {Oliviero}},\ and\ \bibinfo {author} {\bibfnamefont {A.}~\bibnamefont {Hamma}},\ }\bibfield  {title} {\bibinfo {title} {Stabilizer r\'enyi entropy},\ }\href {https://doi.org/10.1103/PhysRevLett.128.050402} {\bibfield  {journal} {\bibinfo  {journal} {Phys. Rev. Lett.}\ }\textbf {\bibinfo {volume} {128}},\ \bibinfo {pages} {050402} (\bibinfo {year} {2022})}\BibitemShut {NoStop}%
\bibitem [{\citenamefont {Haug}\ and\ \citenamefont {Piroli}(2023)}]{Haug2023stabilizerentropies}%
  \BibitemOpen
  \bibfield  {author} {\bibinfo {author} {\bibfnamefont {T.}~\bibnamefont {Haug}}\ and\ \bibinfo {author} {\bibfnamefont {L.}~\bibnamefont {Piroli}},\ }\bibfield  {title} {\bibinfo {title} {Stabilizer entropies and nonstabilizerness monotones},\ }\href {https://doi.org/10.22331/q-2023-08-28-1092} {\bibfield  {journal} {\bibinfo  {journal} {{Quantum}}\ }\textbf {\bibinfo {volume} {7}},\ \bibinfo {pages} {1092} (\bibinfo {year} {2023})}\BibitemShut {NoStop}%
\bibitem [{\citenamefont {Kawai}\ \emph {et~al.}(1986)\citenamefont {Kawai}, \citenamefont {Lewellen},\ and\ \citenamefont {Tye}}]{Kawai:1985xq}%
  \BibitemOpen
  \bibfield  {author} {\bibinfo {author} {\bibfnamefont {H.}~\bibnamefont {Kawai}}, \bibinfo {author} {\bibfnamefont {D.~C.}\ \bibnamefont {Lewellen}},\ and\ \bibinfo {author} {\bibfnamefont {S.~H.~H.}\ \bibnamefont {Tye}},\ }\bibfield  {title} {\bibinfo {title} {{A Relation Between Tree Amplitudes of Closed and Open Strings}},\ }\href {https://doi.org/10.1016/0550-3213(86)90362-7} {\bibfield  {journal} {\bibinfo  {journal} {Nucl. Phys. B}\ }\textbf {\bibinfo {volume} {269}},\ \bibinfo {pages} {1} (\bibinfo {year} {1986})}\BibitemShut {NoStop}%
\bibitem [{\citenamefont {Liu}\ \emph {et~al.}(2026)\citenamefont {Liu}, \citenamefont {Low},\ and\ \citenamefont {Yin}}]{liu2025maximal}%
  \BibitemOpen
  \bibfield  {author} {\bibinfo {author} {\bibfnamefont {Q.}~\bibnamefont {Liu}}, \bibinfo {author} {\bibfnamefont {I.}~\bibnamefont {Low}},\ and\ \bibinfo {author} {\bibfnamefont {Z.}~\bibnamefont {Yin}},\ }\bibfield  {title} {\bibinfo {title} {Maximal magic for two-qubit states},\ }\href {https://doi.org/10.1088/2058-9565/ae3028} {\bibfield  {journal} {\bibinfo  {journal} {Quantum Science and Technology}\ }\textbf {\bibinfo {volume} {11}},\ \bibinfo {pages} {015035} (\bibinfo {year} {2026})}\BibitemShut {NoStop}%
\bibitem [{\citenamefont {Ac\'{\i}n}\ \emph {et~al.}(2001)\citenamefont {Ac\'{\i}n}, \citenamefont {Latorre},\ and\ \citenamefont {Pascual}}]{Acin2001three-party}%
  \BibitemOpen
  \bibfield  {author} {\bibinfo {author} {\bibfnamefont {A.}~\bibnamefont {Ac\'{\i}n}}, \bibinfo {author} {\bibfnamefont {J.~I.}\ \bibnamefont {Latorre}},\ and\ \bibinfo {author} {\bibfnamefont {P.}~\bibnamefont {Pascual}},\ }\bibfield  {title} {\bibinfo {title} {Three-party entanglement from positronium},\ }\href {https://doi.org/10.1103/PhysRevA.63.042107} {\bibfield  {journal} {\bibinfo  {journal} {Phys. Rev. A}\ }\textbf {\bibinfo {volume} {63}},\ \bibinfo {pages} {042107} (\bibinfo {year} {2001})}\BibitemShut {NoStop}%
\bibitem [{\citenamefont {Falc\~ao}\ \emph {et~al.}(2025)\citenamefont {Falc\~ao}, \citenamefont {Tarabunga}, \citenamefont {Frau}, \citenamefont {Tirrito}, \citenamefont {Zakrzewski},\ and\ \citenamefont {Dalmonte}}]{PhysRevB.111.L081102}%
  \BibitemOpen
  \bibfield  {author} {\bibinfo {author} {\bibfnamefont {P.~R.~N.}\ \bibnamefont {Falc\~ao}}, \bibinfo {author} {\bibfnamefont {P.~S.}\ \bibnamefont {Tarabunga}}, \bibinfo {author} {\bibfnamefont {M.}~\bibnamefont {Frau}}, \bibinfo {author} {\bibfnamefont {E.}~\bibnamefont {Tirrito}}, \bibinfo {author} {\bibfnamefont {J.}~\bibnamefont {Zakrzewski}},\ and\ \bibinfo {author} {\bibfnamefont {M.}~\bibnamefont {Dalmonte}},\ }\bibfield  {title} {\bibinfo {title} {Nonstabilizerness in u(1) lattice gauge theory},\ }\href {https://doi.org/10.1103/PhysRevB.111.L081102} {\bibfield  {journal} {\bibinfo  {journal} {Phys. Rev. B}\ }\textbf {\bibinfo {volume} {111}},\ \bibinfo {pages} {L081102} (\bibinfo {year} {2025})}\BibitemShut {NoStop}%
\bibitem [{\citenamefont {Latosh}(2022)}]{Latosh:2022ydd}%
  \BibitemOpen
  \bibfield  {author} {\bibinfo {author} {\bibfnamefont {B.}~\bibnamefont {Latosh}},\ }\bibfield  {title} {\bibinfo {title} {{FeynGrav: FeynCalc extension for gravity amplitudes}},\ }\href {https://doi.org/10.1088/1361-6382/ac7e15} {\bibfield  {journal} {\bibinfo  {journal} {Class. Quant. Grav.}\ }\textbf {\bibinfo {volume} {39}},\ \bibinfo {pages} {165006} (\bibinfo {year} {2022})},\ \Eprint {https://arxiv.org/abs/2201.06812} {arXiv:2201.06812 [hep-th]} \BibitemShut {NoStop}%
\bibitem [{\citenamefont {Sannan}(1986)}]{Sannan:1986tz}%
  \BibitemOpen
  \bibfield  {author} {\bibinfo {author} {\bibfnamefont {S.}~\bibnamefont {Sannan}},\ }\bibfield  {title} {\bibinfo {title} {{Gravity as the Limit of the Type {II} Superstring Theory}},\ }\href {https://doi.org/10.1103/PhysRevD.34.1749} {\bibfield  {journal} {\bibinfo  {journal} {Phys. Rev. D}\ }\textbf {\bibinfo {volume} {34}},\ \bibinfo {pages} {1749} (\bibinfo {year} {1986})}\BibitemShut {NoStop}%
\end{thebibliography}%

\end{document}